\documentclass[%
reprint,
superscriptaddress,
amsmath,
amssymb,
aps,
prb,
floatfix,
]{revtex4-2}

\usepackage{graphicx}
\usepackage{subfigure}
\usepackage{bm}
\usepackage{txfonts}
\usepackage[colorlinks=true,citecolor=blue,urlcolor = blue,linkcolor= blue]{hyperref}
\usepackage{epstopdf}
\usepackage{color,soul}
\usepackage[toc,page]{appendix}

\makeatletter
\AtBeginDocument{\@ifpackageloaded{natis}{\ifNAT@numbers\if@filesw\immediate\write\@auxout{\string\global\string\NAT@numberstrue}\fi\fi}{}}
\makeatother

\begin{document}

\title{Hidden spatiotemporal sequence in transition to shear band in amorphous solids}

\author{Zeng-Yu Yang}
\affiliation{State Key Laboratory of Nonlinear Mechanics, Institute of Mechanics, Chinese Academy of Sciences, Beijing 100190, China}
\affiliation{School of Engineering Science, University of Chinese Academy of Sciences, Beijing 100049, China}

\author{Yun-Jiang Wang}
\affiliation{State Key Laboratory of Nonlinear Mechanics, Institute of Mechanics, Chinese Academy of Sciences, Beijing 100190, China}
\affiliation{School of Engineering Science, University of Chinese Academy of Sciences, Beijing 100049, China}

\author{Lan-Hong Dai}
\email{lhdai@lnm.imech.ac.cn}
\affiliation{State Key Laboratory of Nonlinear Mechanics, Institute of Mechanics, Chinese Academy of Sciences, Beijing 100190, China}
\affiliation{School of Engineering Science, University of Chinese Academy of Sciences, Beijing 100049, China}
\affiliation{School of Future Technology, University of Chinese Academy of Sciences, Beijing 100049, China}

\date{\today}

\begin{abstract}

Localization of plastic flow into narrow shear band is a fundamental and ubiquitous non-equilibrium phenomenon in amorphous solids. Because of the intrinsic entangling of three elementary local atomic motions: shear, dilatation and rotation, the precise physical process of shear band emergence is still an enigma. Here, to unveil this mystery, we formulate for the first time a theoretical protocol covering both affine and non-affine components of deformation, to decode these three highly entangled local atomic-scale events. In contrast to the broad concept of shear transformation zone, the plastic behavior can be demonstrated comprehensively as the operative manipulation of more exact shear-dominated zones (SDZs), dilatation-dominated zones (DDZs) and rotation-dominated zones (RDZs). Their spatiotemporal evolution exhibits a novel transition from synchronous motion to separate distribution at the onset of shear band, which is in striking resemblance with the transition from laminar flow to turbulent flow in flow dynamics. The hidden mechanism is then revealed with the help of extreme value theory (EVT) and percolation analysis. Numerical evidence from EVT indicates that dilatation is the dominant mode at the embryo of initial plastic units, as evidenced through larger degree of dilatation localization over shear and rotation. The percolation analysis points towards the critical power-law scaling nature at the transition from stochastic activation to percolation of plastic regions. Then the comprehensive pictures underlying shear banding emergence is uncovered. Firstly, dilatation triggers initial shear and rotation in soft regions, leading to embryo of initial flow units, which is followed by the inhomogeneous turbulent-like pattern manifesting as the secondary activation of rotation in neighboring hard material. Such rotation activation contributes to further perturbation in these regions, and ultimately, leads to percolation transition and shear band formation. Our findings also reinforce that the discussion of plastic behavior in disordered materials must take into account both affine and non-affine component deformation.

\end{abstract}

\maketitle

\section{Introduction}
Transition from smooth and homogeneous plastic flow to unstable and inhomogeneous flow in the form of thin highly localized shear band is a ubiquitous and fundamental nonequilibrium phenomenon in condensed matter ranging from metals, polymers, glasses, colloids, and granular media, etc. \cite{Greer2013,Schuh2007,Falk2011,Gourlay2007,Schall2010,Furukawa2009,Bai1992,Wisitsorasaka2017,Wang2019,Yan2021}. For metallic glasses (MGs), shear band is their dominant deformation mode at room temperature, which can induce catastrophic fracture with very limited ductility, impeding further engineering applications \cite{Spaepen1977,Langer2001,Manning2009,Egami2013,Greer2013,DAI2012311}. Therefore, it is extremely important to explore the physical origin of shear band in MGs. However, in contrast to their crystalline counterparts, whose plastic flow is given by the well-defined topological defects in the periodic lattice such as dislocations, plastic deformation mechanisms of MGs are less well understood due to their inherent structural disorder \cite{Cheng2011,Fan2014,Ye2010,Tong2018}. To date, numerous theoretical models have been proposed to describe the local flow event in MGs, such as free volume model \cite{Spaepen1977,Spaepen2006,Wang2017}, shear transformation zone (STZ) \cite{Argon1979,Falk1998,Lemaitre2002}, cooperative shear model \cite{Johnson2005}, flow units \cite{Lu2014,WangZ2019}, soft spot \cite{Ding2014}, and tension transformation zone (TTZ) \cite{Jiang2008,Huang2014}, among which the shear transformation zone, involving local rearrangement of a small cluster of atoms, is generally accepted as the elementary process of plastic deformation in MGs \cite{Pan2008,Manning2007}. In this connection, considerably experimental \cite{Lewandowski2006,Schmidt2015,Shen2018} and theoretical attempts \cite{Huang2002,Dai2005,Jiang2009,Han2009} have been made to give the bridge between STZ's motion and shear banding emergence. Later simulations indicates that the continued propagation of shear strain occurs by a process of self-assembly: the operation of one STZ creates a localized distortion of the surrounding material, and triggers the autocatalytic formation of large planar bands of STZs, causing shear band formation \cite{Shimizu2006,Cao2009,Li2007}. This is so far most widely accepted scenario of shear banding.

In spite of this, further experimental and simulated evidences were collected later on, revealing the free-volume annihilation and creation process \cite{Klaumunzer2011,Schmidt2015,Zeng2018} as well as vortex-like motion \cite{Maloney2006,Hassani2019,Sopu2017} in MGs, supporting the general consensus that shear is not necessarily the only deformation mode that accommodates the local atomic rearrangement. Actually, all of shear, dilatation and rotation are intrinsic natures of local excitations in MGs, which cannot be fully explained via the classical STZ model. It is the inherent coupling of these deformation events that hinders establishment of an intuitive picture of shear banding emergence in metallic glasses. In this connection, Şopu et al. \cite{Sopu2017} recently proposed the STZ-vortex two-unit model to address the issue of sequential activation of STZs along the direction of shear band. The STZ-vortex model proposes the mechanism of STZ coalesce with critical role of rotation events as catalyst, which mediates distortion between two adjacent STZs \cite{Sopu2017}. Despite its notable progress in providing atomistic description of shear banding, the STZ-vortex model has not totally decoded the entangled atomic motions hidden in the non-affine deformation section. In Ref \cite{Sopu2017}, the non-affine displacement \cite{Falk1998} has been calculated and discussed as a whole, however, it was not decoupled into shear, dilatation and rotation events. The decoupling method in the STZ-vortex framework is based on the purely affine part of local displacement field -- local deformation gradient tensor \cite{Shimizu2007}. It should be noted that only affine deformation may miss the meaningful mechanism hidden in the non-affine components. The latter is also strongly correlated with the inhomogeneous plastic deformation in the generic disordered materials \cite{Falk1998,Richard2020,Zhang2021}. Therefore, to thoroughly unveil the shear banding mechanism at atomic scale, a decoupling framework considering both of affine and non-affine deformation is urgently necessary.

To this end, we propose a theoretical framework that can incorporate both affine and non-affine parts of deformation by extensively combining the first and second order of displacement gradient tensor. As we will later demonstrate via a testing simulated Cu$_{50}$Zr$_{50}$ MG, information from single affine or non-affine input can only partially capture the nature of plastic deformation. The addition of second order term significantly enhances the capacity of theoretical model in mapping the real displacement field. On the basis of this powerful framework, we decouple the highly entangled shear, dilatation and rotation events and thus give the precise identification of these three deformation units. As a step beyond the usual broad concept of STZ percolation, the comprehensive and clear atomic-scale physical process of shear banding is then rationalized via the complicated and obscure interplays between the versatile shear, dilatation and rotation events. Firstly, shear, dilatation and rotation are strongly correlated with each other, among which dilatation plays a dominating role in the embryo of initial plastic events in liquid-like regions. Then, with strain goes on, secondary rotation in adjacent solid-like materials is significantly activated, causing the transition from homogeneous laminar-like motion to inhomogeneous turbulent-like flow. Such rotation activation contributes to further softening and perturbation in these regions, and ultimately, leads to the connection of co-existing localized plastic regions characterized as shear band emergence. The connection phenomenon exhibits a percolation transition with power-law scaling nature that is in consistent with classical percolation theory. Our findings shed new light on fundamental understanding of shear band formation in metallic glasses and other amorphous materials.

\section{Simulation details}
The molecular dynamic simulations were performed using the LAMMPS code \cite{Plimpton1995}, with the embedded-atom method (EAM) potential \cite{Cheng2009} being adopted to describe the atomic interactions for the Cu$_{50}$Zr$_{50}$ MG. A small sample containing 13500 atoms was melted from its crystalline phase from 100 K to 2100 K, and then being equilibrated for 500 ps at 2100 K before quenched to a glassy state (100 K) at a cooling rate of 0.02 K/ps. A larger model system to be deformed, containing $\sim$ 660 000 atoms, with dimensions of $37.3 \times 6.2 \times 61.7$ nm$^3$ in the $x$, $y$, $z$ directions, respectively, was then produced by the replication of the initial glass configuration. The large system was further annealed for 500 ps at 800 K to reduce the artificial boundary effect of multiplication. Periodic boundary conditions (PBCs) were adopted for each direction during the sample preparation process. To ensure that there would be only a single shear band during the deformation, a small notch was created in the rectangular sample to yield a stress concentration on the notch and nucleation of the shear band. Uniaxial loading, with a constant strain rate of $4 \times 10^7$ s$^{-1}$, was employed to the notched sample along the $z$ direction at a low temperature of 100 K. While PBCs were imposed along the $y$ and $z$ directions, the free surface condition was applied in the $x$ direction to enable the occurrence of the shear offset on the free surfaces during the deformation process. The pressure and temperature were controlled using isothermal-isobaric ($N$ atom number, $P$ pressure, and $T$ temperature) ensembles \cite{Parrinello1981} and Nose-Hoover thermostat \cite{Nose1984,Hoover1985} for both the sample preparation and uniaxial loading process. The MD time step was 0.001 ps.

\section{Theoretical framework}
\subsection{Mapping the deformation field}

\begin{figure*}
  \centering
  \includegraphics[width=0.8\textwidth]{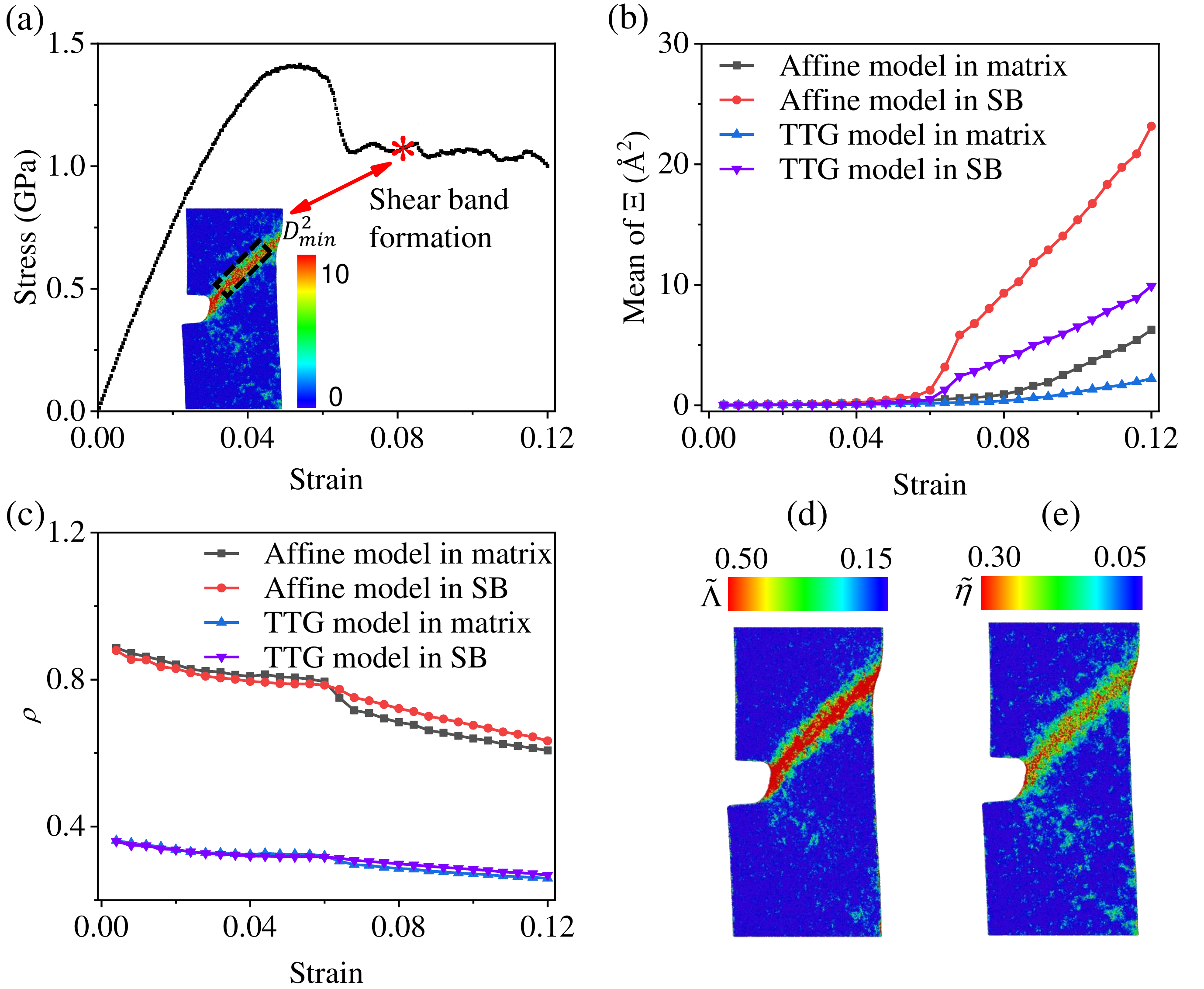}
  \caption{The powerful effect of TTG model in describing the deformation field of metallic glass. (a) The stress-strain curve and the distribution of $D_{\min }^2$ after yielding with the dotted boxes locating the position of the mature shear band. (b) Evolution of residual displacement lost by the affine framework and TTG model respectively. (c) Evolution of the relative difference of displacements between the whole deformation field and predictive parts described by affine model and TTG model respectively. (d) Distribution of atomic effective strain at strain of 0.08. (e) Distribution of atomic effective strain gradient at strain of 0.08.}\label{Fig:1}
\end{figure*}

Based on the continuum mechanics, the local transformation relation between reference and current configurations during a time interval $\Delta{t}$ can be described as the form of Taylor expansion:
\begin{equation}\label{eq:1}
  {{\bf{d}}^{ij}}\left( {t + \Delta t} \right) = {{\bf{F}}^i}{{\bf{d}}^{ij}}\left( t \right) + \frac{1}{2}{{\bm{\eta }}^i}{{\bf{d}}^{ij}}{\left( t \right)^2} + R\left( {{{\bf{d}}^{ij}}{{\left( t \right)}^3}} \right).
\end{equation}
Here, vectors and tensors are given in bold font. The superscript $i$ and $j$ are used to distinguish different atoms. ${{\bf{d}}^{ij}}\left( t \right)$ is the center to center position vector between central atom $i$ and its neighboring $j$th atom. Thus, $t=0$ characterizes the initial configuration before any applied strain. Amongst the Taylor expansion, the first term ${{\bf{F}}^i}{{\bf{d}}^{ij}}\left( t \right)$ denotes the linear relation or affine part of the local displacement field around central atom $i$. ${{\bf{F}}^i}$ is the deformation gradient tensor with the definition of ${{\bf{F}}^i} = {{\bf{H}}^i} + {\bf{I}}$, where the displacement gradient tensor ${{\bf{H}}^i}$ characterizes the spatial gradient of displacement field, and $\mathbf{I}$ is the identity tensor representing the rigid translation. In terms of this affine part, Shimizu et al \cite{Shimizu2007} identified the local Lagrangian strain tensor ${{\bm{\varepsilon }}^i} = \frac{1}{2}\left[ {{{\bf{F}}^i}{{\left( {{{\bf{F}}^i}} \right)}^T} - {\bf{I}}} \right]$ and proposed a local von Mises strain as $\varepsilon _{{\rm{Mises}}}^i = \sqrt {\varepsilon _{xy}^2 + \varepsilon _{yz}^2 + \varepsilon _{zx}^2 + \frac{{{{\left( {{\varepsilon _{xx}} - {\varepsilon _{yy}}} \right)}^2} + {{\left( {{\varepsilon _{xx}} - {\varepsilon _{zz}}} \right)}^2} + {{\left( {{\varepsilon _{zz}} - {\varepsilon _{yy}}} \right)}^2}}}{6}}$. Then, the second and higher order terms of Eq. (\ref{eq:1}), $\frac{1}{2}{{\bm{\eta }}^i}{{\bf{d}}^{ij}}{\left( t \right)^2} + R\left( {{{\bf{d}}^{ij}}{{\left( t \right)}^3}} \right)$, construct the non-affine part of displacement field. Here, ${\bm{\eta} ^i}$ is the second-order displacement gradient tensor i.e., strain gradient tensor. To characterizes departures from the local affine deformation, Falk and Langer \cite{Falk1998} proposed the widely used non-affine squared displacement $D_{\min }^2 = \sum\limits_j {{{\left[ {{{\bf{d}}^{ij}}\left( {t + \Delta t} \right) - {{\bf{F}}^i}{{\bf{d}}^{ij}}\left( t \right)} \right]}^2}}$.

Both of non-affine displacement and local von Mises strain are good measures of deformation and have raised broad interest and impact in the field of glassy physics \cite{Peng2011,Cubuk2017,Cao2014,Cao2017}. However, these two quantities actually characterize distinct parts of deformation in terms of their definitions. On the one hand, the local von Mises strain is the output of local linear strain field. Thus, it depicts the affine part of deformation. On the other hand, $D_{\rm{min}}^2$ quantifies departures from the local linear strain field, and thus characterizes the non-affine part. Both of these quantities have achieved great success in mapping deformation of disordered materials \cite{Wang2018,Hufnagel2016,Sha2015}. However, it is still a challenge to distinguish and characterize the rotation, dilatation, and shear events via these theoretical models, as rotation is mostly related to rigid motion which is affine while dilatation as well as shear are dominated by distortion which is non-affine. To settle this problem, a further theoretical framework containing both affine and non-affine deformation is urgently needed, which then drives the birth of our proposed two-term gradient (TTG) model. On the basis of linear affine formalism, the second term of Eq. (\ref{eq:1}) which incorporates strain gradient tensor, is supplemented to recognize the non-affine part. Thus the displacement field is assumed to vary as:
\begin{equation}\label{eq:2}
  {{\bf{d}}^{ij}}\left( {t + \Delta t} \right) \approx {{\bf{F}}^i}{{\bf{d}}^{ij}}\left( t \right) + \frac{1}{2}{{\bm{\eta }}^i}{{\bf{d}}^{ij}}{\left( t \right)^2}.
\end{equation}
Then, ${{\bf{F}}^i}$ and ${\bm{\eta} ^i}$ are calculated by minimizing the mean-squared difference between the actual displacement field and that demonstrated by ${{\bf{F}}^i}$ and ${\bm{\eta} ^i}$:
\begin{equation}\label{eq:3}
 {\Xi ^i} = \frac{1}{{{N^i}}}\sum\limits_j {{{\left[ {{{\bf{d}}^{ij}}\left( {t + \Delta t} \right) - {{\bf{F}}^i}{{\bf{d}}^{ij}}\left( t \right) - \frac{1}{2}{{\bm{\eta }}^i}{{\bf{d}}^{ij}}{{\left( t \right)}^2}} \right]}^2}}.
\end{equation}
Here, $N^i$ is the number of neighboring atoms around central atom $i$. The minimum value of $\Xi ^i$ is then used to quantify the capacity of our TTG model in mapping the real displacement field.

In the present work, we employ molecular dynamics (MD) simulations in a prototypical binary Cu$_{50}$Zr$_{50}$ glass as a computational microscope to test and demonstrate the TTG model. Figure \ref{Fig:1} shows the strong power of TTG model in describing the real displacement field. Firstly, a typical stress-strain curve is presented in Fig. \ref{Fig:1}a, from which it is clear that deformation localization appears at a macroscopic strain of 0.06, while the formation of a mature shear band falls behind at a larger strain (0.08), as marked with an asterisk in the plot. In the inset of Fig. \ref{Fig:1}a, the spatial distribution of $D_{\rm{min}}^2$ at strain of 0.08 is shown and one can visualize shear banding pattern outlined by the black box. In Fig. \ref{Fig:1}b, we plot the evolution of $\Xi ^i$ for atoms in the shear band as well as those taken from the matrix. For comparison, the departure of the displacement field assumed by affine model from the real one is also presented. Here, the behavior of deformation localization can be observed at the strain 0.06 which meets the stress-strain curve well. More interesting, the difference between the actual displacement field and TTG model is much less than that of the affine model. It gives the direct evidence for the strong enhancement of mapping power of TTG model over the linear affine model. It reinforces the importance of incorporating the strain gradient term. To further quantify how much the deformation information is omitted by the TTG model comparing to the whole displacement field, we propose the measure:
\begin{widetext}
\begin{equation}\label{eq:4}
{\rho _{{\rm{TTG}}}} = \frac{1}{N}\sum\nolimits_i {\frac{{\left| {{{\sum\nolimits_j {\left[ {{{\bf{F}}^i}{{\bf{d}}^{ij}}\left( t \right) + \frac{1}{2}{{\bm{\eta }}^i}{{\bf{d}}^{ij}}{{\left( t \right)}^2} - {{\bf{d}}^{ij}}\left( t \right)} \right]} }^2} - \sum\nolimits_j {{{\left[ {{{\bf{d}}^{ij}}\left( {t + \Delta t} \right) - {{\bf{d}}^{ij}}\left( t \right)} \right]}^2}} } \right|}}{{\sum\nolimits_j {{{\left[ {{{\bf{d}}^{ij}}\left( {t + \Delta t} \right) - {{\bf{d}}^{ij}}\left( t \right)} \right]}^2}} }}},
\end{equation}
\end{widetext}
which is the ratio of the omitted displacements to the actual ones. $N$ is the total number of atoms in the sample. For comparison, a similar definition is used for affine model:
\begin{widetext}
\begin{equation}\label{eq:5}
{\rho _{{\rm{affine}}}} = \frac{1}{N}\sum\nolimits_i {\frac{{\left| {{{\sum\nolimits_j {\left[ {{{\bf{F}}^i}{{\bf{d}}^{ij}}\left( t \right) - {{\bf{d}}^{ij}}\left( t \right)} \right]} }^2} - \sum\nolimits_j {{{\left[ {{{\bf{d}}^{ij}}\left( {t + \Delta t} \right) - {{\bf{d}}^{ij}}\left( t \right)} \right]}^2}} } \right|}}{{\sum\nolimits_j {{{\left[ {{{\bf{d}}^{ij}}\left( {t + \Delta t} \right) - {{\bf{d}}^{ij}}\left( t \right)} \right]}^2}} }}}.
\end{equation}
\end{widetext}
The ratio $\rho$ for both of affine and TTG model as a function of macroscopic strain are shown in Fig. \ref{Fig:1}c. It provides the direct evidence that the present TTG model with integration of strain gradient effect leads to much higher mapping capability. TTG model contains most of the deformation information whatever the atom is in shear band or matrix. This implies that deformation in metallic glasses is more related to the non-affine components, of which TTG model contains the crucial part -- the second order term. Such success of TTG model confirms the validity and reliability of our decoupling results and conclusions below.

\begin{figure}
  \centering
  \includegraphics[width=0.5\textwidth]{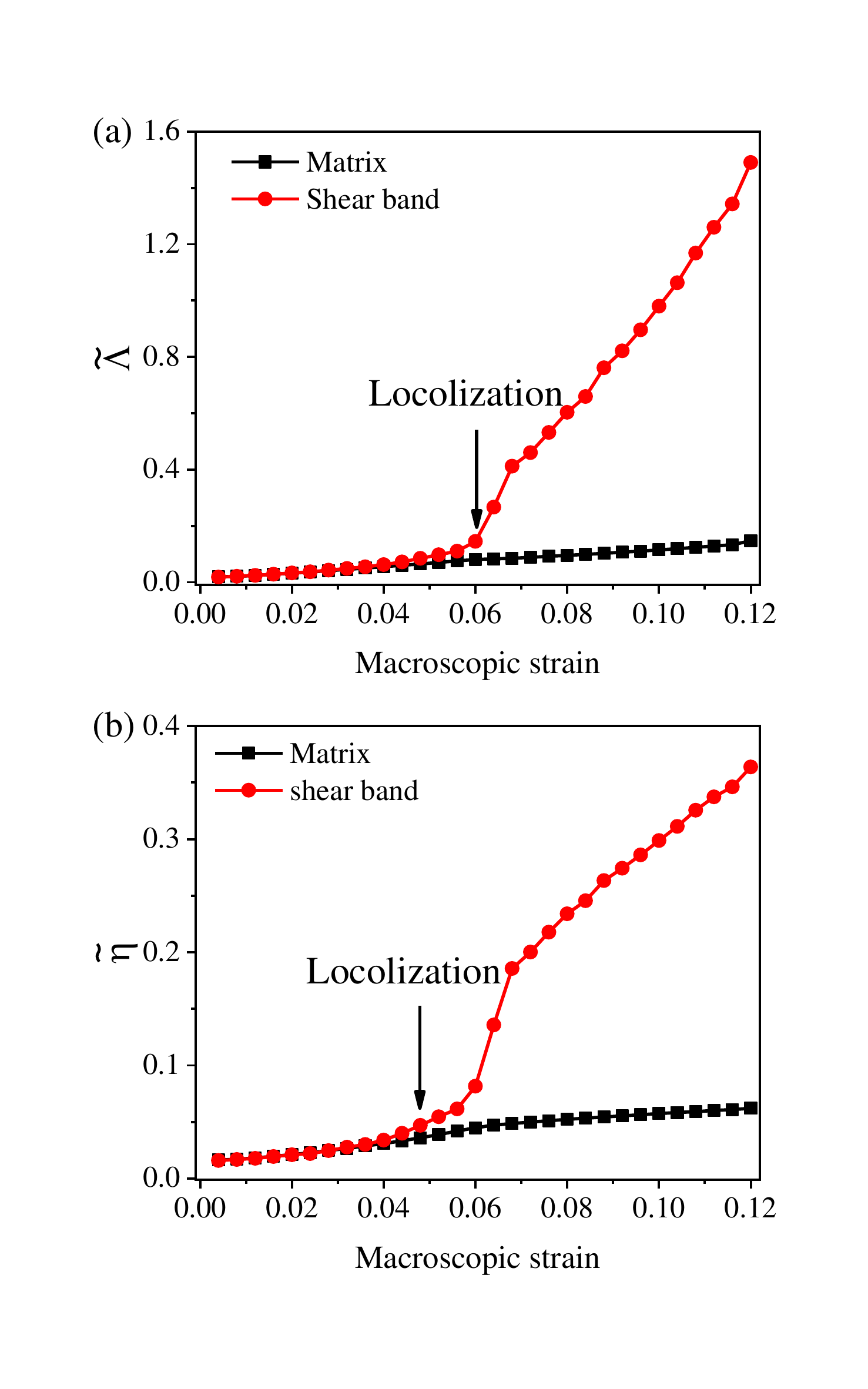}
  \caption{Evolution of (a) effective atomic strain and (b) effective strain gradient during shear banding. The results are calculated for atoms in shear band and matrix, respectively. Strain gradient effect is apparent in shear band during deformation. Much of the large strain gradient locates inside of shear band. The black arrows point out the critical points when the localization of strain or strain gradient occurs.}\label{Fig:2}
\end{figure}

Before we present the decoupling method in next subsection, it is useful to discuss the critical role of strain gradient in shear banding. Firstly, to quantify plastic deformation at the atomic scale, we introduce the atomic local strain for each atom based on the TTG model. Following the general expression given by Gao et al \cite{Gao1999}, the microscale strain field incorporating strain gradient is defined as $\Lambda _{mn}^i = \varepsilon _{mn}^i + \frac{1}{{{N^i}}}\sum\limits_j {\left[ {\frac{1}{2}\left( {\eta _{kmn}^i + \eta _{knm}^i} \right)d_k^{ij}(t)} \right]}$, where the subscripts $k$, $m$ and $n$ indicate the Cartesian components, e.g. $x$ or $y$ or $z$ and the superscripts $i$, $j$ are particle indices. Here, $d_{k}^{ij}(t)$ is the $k$th component of the position of the $j$th neighboring atom relative to center atom $i$ in the reference configuration, and $N^i$ is the number of neighboring atoms around atom $i$. The atomic strain filed $\Lambda_{mn}^i$ is thus related to the strain tensor $\varepsilon_{mn}^i$ and strain gradient tensor $\eta_{kmn}^i$. Following the classical plasticity theory, the effective strain as the form of scalar is then introduced as $\tilde \Lambda ^i = \sqrt {\frac{2}{3}{\Lambda _{mn}^i}{\Lambda _{mn}^i}}$. The spatial distribution of $\tilde \Lambda ^i$ at strain of 0.08 is shown in Fig. \ref{Fig:1}d, like Falk and Langer's $D_{\rm{min}}^2$ \cite{Falk1998}, $\tilde \Lambda ^i$ is a good measure of local inelastic deformation. More projected strain fields color coded by $\tilde \Lambda ^i$ at various applied strains are shown in Fig. S1 of Supplemental Material (SM) \cite{sm}. Then the atomic effective strain gradient as the form of scalar is introduced as $\tilde \eta ^i = \sqrt {\frac{1}{4}{\eta _{kmn}^i}{\eta _{kmn}^i}}$, which is similar to that of the effective strain \cite{Gao1999}. The projected strain gradient field at strain of 0.08 is plotted in Fig. \ref{Fig:1}e. Here, we can find good correspondence between the strain gradient field and the distribution of deformation. More snapshots color coded by strain gradient are given in Fig. S2 \cite{sm}. Local regions with accumulated strain penetrate into each other overlaps locations where strain gradient effect is apparent. It indicates that the strain gradient effect does play an important role in shear banding. To clarify the critical role played by strain gradient during shear banding process, the statistic effective strain and strain gradient as functions of macroscopic strain are given in Fig. \ref{Fig:2}. The results are calculated for atoms in shear band and matrix, respectively. It shows that the deviation between shear band and matrix for effective strain gradient is earlier than that of effective strain. It is the direct evidence that strain gradient takes precedence over strain localization and promotes the formation of shear band. This result is consistent with our previous works \cite{Tian2017,Dai2004,Liu2020}, which demonstrate a self-feedback mechanism that high strain gradient, acting as the driving force, will induce inhomogeneous energy dissipation and thus aggravates deformation localization.

\subsection{Decoupling the entangled shear, dilatation and rotation events}

\begin{figure*}
  \centering
  \includegraphics[width=0.9\textwidth]{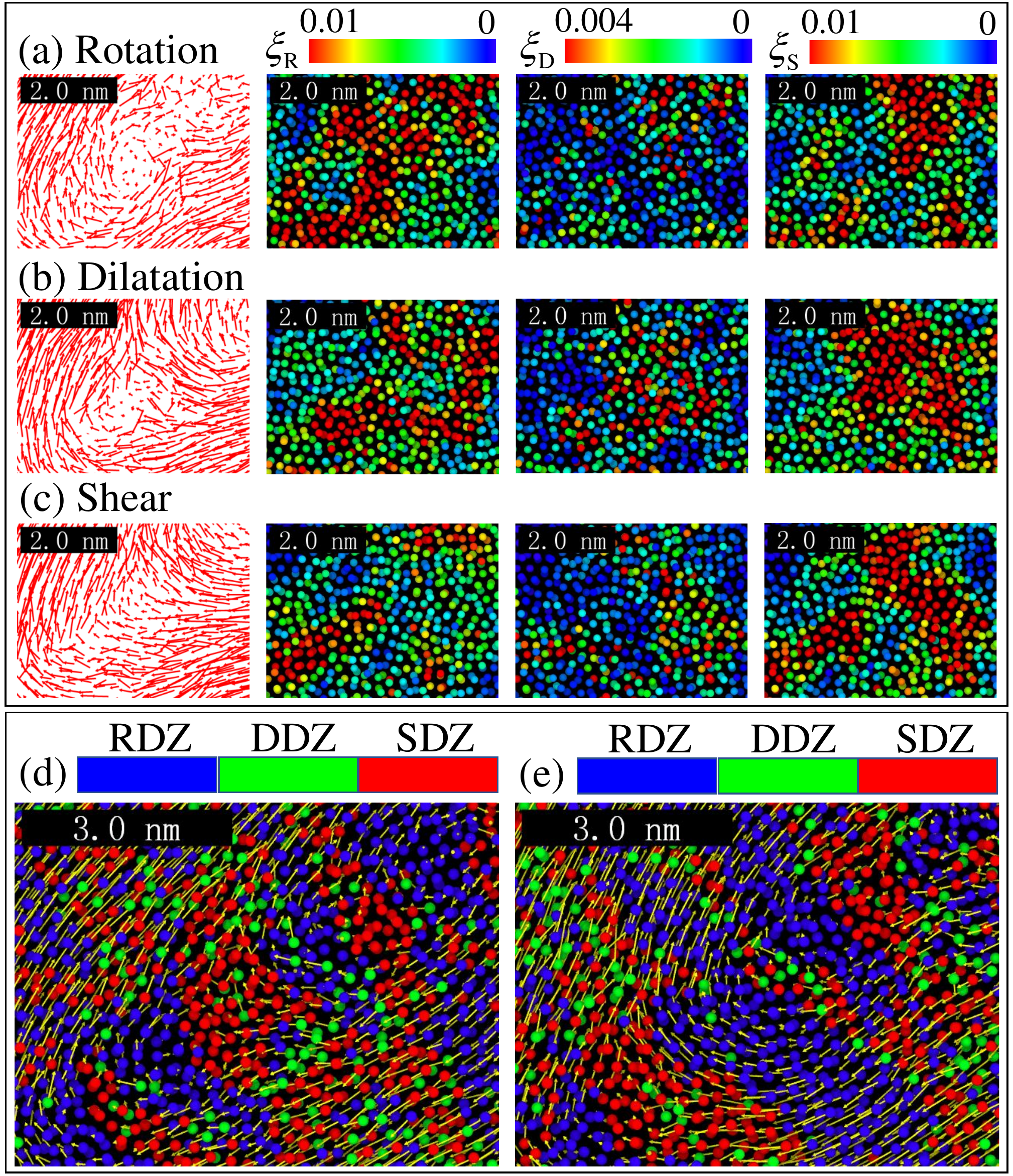}
  \caption{Representative illustrations of (a) rotation, (b) dilatation, and (c) shear events in the shear band described by the displacement vectors, magnitudes of ${\xi _{\rm{R}}}$, ${\xi _{\rm{D}}}$, ${\xi _{\rm{S}}}$ (color code), respectively. (d) Representative SDZ and DDZ where atoms mainly undergo shear and dilatation transformation, as denoted by the red and green atoms. (e) Representative RDZ where atoms mainly undergo rotation transformation as denoted by blue atoms. The yellow arrows represent the displacement vectors, and the atoms are colored according to their RDZ, DDZ, and SDZ labels, respectively. All of snapshots are taken with a slice of 5 {\AA} perpendicular to the paper plane.}\label{Fig:3}
\end{figure*}

Next, we explain the proposed method to decouple the entangled shear, dilatation and rotation events. Here, the index notation of tensor analysis is used to demonstrate the details of the decoupling method. In the matter of the affine part, as mentioned above, the deformation gradient tensor $F_{mn}^i = H_{mn}^i + I_{mn}$. As $I_{mn}$ is the identity tensor which represents the rigid translation and contains no effective deformation, we just need to consider the decoupling of displacement gradient tensor, $H_{mn}^i$, which can be rearranged into three parts, i.e.,
\begin{equation}\label{eq:6}
\begin{aligned}
  {H_{mn}^i} &= \frac{1}{2}\left( {{H_{mn}^i} - {H_{nm}^i}} \right) \\
   &+ \left[ {\frac{1}{2}\left( {{H_{mn}^i} + {H_{nm}^i}} \right) - \frac{1}{3}\left( {{H_{mn}^i}{I_{mn}}} \right)} \right] \\
   &+ \frac{1}{3}\left( {{H_{mn}^i}{I_{mn}}} \right),
\end{aligned}
\end{equation}
where the term $\frac{1}{2}(H_{mn}^i - H_{nm}^i)$, $\left[ {\frac{1}{2}\left( {{H_{mn}^i} + {H_{nm}^i}} \right) - \frac{1}{3}\left( {{H_{mn}^i}{I_{mn}}} \right)} \right]$ and $\frac{1}{3}(H_{mn}^iI_{mn})$ on the right-hand side characterizes the rotation, shear and dilatation component respectively. All of these are related to affine deformation. That is
\begin{equation}\label{eq:7}
\begin{aligned}
  R_{mn}^{H,i} &= \frac{1}{2}\left( {{H_{mn}^i} - {H_{nm}^i}} \right) \\
  S_{mn}^{H,i} &= \left[ {\frac{1}{2}\left( {{H_{mn}^i} + {H_{nm}^i}} \right) - \frac{1}{3}\left( {{H_{mn}^i}{I_{mn}}} \right)} \right] \\
  D_{mn}^{H,i} &= \frac{1}{3}\left( {{H_{mn}^i}{I_{mn}}} \right)
\end{aligned}
\end{equation}
As for the decoupling method related to the strain gradient tensor, following the notation of Fleck and Hutchinson \cite{Fleck1997}, $D_{kmn}^{\eta ,i} = \frac{1}{8}\left( {{\delta _{kn}}\eta _{mll}^i + {\delta _{mn}}\eta _{kll}^i} \right)$ denotes the hydrostatic i.e., dilatation part of $\frac{1}{2}{\bm{\eta} ^i}$. Here, $\delta_{kn}$ is the Kronecker symbol. Then the deviatoric component follows as $\frac{1}{2}\eta '^i_{kmn} = \frac{1}{2}\eta _{kmn}^i - D_{kmn}^{\eta ,i}$, of which the fully symmetric tensor $S_{kmn}^{\eta,i}  = \frac{1}{6}\left( {{{\eta '}^i_{kmn}} + {{\eta '}^i_{mnk}} + {{\eta '}^i_{nkm}}} \right)$ is introduced to denote the shear part. Therefore, the remaining part $R_{kmn}^{\eta,i}  = {\eta '^i_{kmn}} - S_{kmn}^{\eta,i}$ can be utilized to specify the rotation related to strain gradient tensor. Based on the decoupling of the affine and non-affine deformation, we can introduce the scalar product of these tensors as:
\begin{equation}\label{eq:8}
\begin{aligned}
{\xi _{\rm{R}}^i} &= \sqrt {R_{mn}^{H,i}R_{mn}^{H,i}}  \cdot \sqrt {R_{kmn}^{\eta,i} R_{kmn}^{\eta,i} } \\
{\xi _{\rm{D}}^i} &= \sqrt {D_{mn}^{H,i}D_{mn}^{H,i}}  \cdot \sqrt {D_{kmn}^{\eta,i} D_{kmn}^{\eta,i} } \\
{\xi _{\rm{S}}^i} &= \sqrt {S_{mn}^{H,i}S_{mn}^{H,i}}  \cdot \sqrt {S_{kmn}^{\eta,i} S_{kmn}^{\eta,i} }
\end{aligned}
\end{equation}
which are called the rotation transformation factor, dilatation transformation factor, and shear transformation factor, respectively. These factors are used to quantitatively describe the rotation, dilatation, and shear transformation events in the model glass. It should be noted that higher $\xi_{\rm{R}}^i$, $\xi_{\rm{D}}^i$, and $\xi_{\rm{S}}^i$ values indicate a more severe level of corresponding local transformation motion. In Fig. \ref{Fig:3}a-c, the representative shear, dilatation and rotation events are revealed by virtue of the atomic displacement vectors. Here, we can carefully observe good spatial correspondence among the magnitudes of $\xi_{\rm{R}}^i$, $\xi_{\rm{D}}^i$, and $\xi_{\rm{S}}^i$ and the exact locations of rotation, dilatation, and shear events. This correspondence indicates that these three transformation factors, $\xi_{\rm{R}}^i$, $\xi_{\rm{D}}^i$, and $\xi_{\rm{S}}^i$ are indeed excellent diagnostics for identifying the local rotation, dilatation and shear events, respectively.

In order to quantitatively distinguish the relative roles of shear, dilatation, and rotation events at atomic-scale, we further introduce three transformation zones, namely, the rotation-dominated zone (RDZ), dilatation-dominated zone (DDZ), and shear-dominated zone (SDZ). First, the participation fraction of the rotation, dilatation, and shear in the deformation of an atom is defined as:
\begin{equation}\label{eq:9}
\begin{aligned}
\chi _{\rm{R}}^i = \frac{{{{\xi _{\rm{R}}^i} \mathord{\left/
 {\vphantom {{\xi _{\rm{R}}^i} {\xi _{\rm{R}}^{\rm{M}}}}} \right.
 \kern-\nulldelimiterspace} {\xi _{\rm{R}}^{\rm{M}}}}}}{{\sqrt {{{\left( {{{\xi _{\rm{R}}^i} \mathord{\left/
 {\vphantom {{\xi _{\rm{R}}^i} {\xi _{\rm{R}}^{\rm{M}}}}} \right.
 \kern-\nulldelimiterspace} {\xi _{\rm{R}}^{\rm{M}}}}} \right)}^2} + {{\left( {{{\xi _{\rm{D}}^i} \mathord{\left/
 {\vphantom {{\xi _{\rm{D}}^i} {\xi _{\rm{D}}^{\rm{M}}}}} \right.
 \kern-\nulldelimiterspace} {\xi _{\rm{D}}^{\rm{M}}}}} \right)}^2} + {{\left( {{{\xi _{\rm{S}}^i} \mathord{\left/
 {\vphantom {{\xi _{\rm{S}}^i} {\xi _{\rm{S}}^{\rm{M}}}}} \right.
 \kern-\nulldelimiterspace} {\xi _{\rm{S}}^{\rm{M}}}}} \right)}^2}} }}\\
\chi _{\rm{D}}^i = \frac{{{{\xi _{\rm{D}}^i} \mathord{\left/
 {\vphantom {{\xi _{\rm{D}}^i} {\xi _{\rm{D}}^{\rm{M}}}}} \right.
 \kern-\nulldelimiterspace} {\xi _{\rm{D}}^{\rm{M}}}}}}{{\sqrt {{{\left( {{{\xi _{\rm{R}}^i} \mathord{\left/
 {\vphantom {{\xi _{\rm{R}}^i} {\xi _{\rm{R}}^{\rm{M}}}}} \right.
 \kern-\nulldelimiterspace} {\xi _{\rm{R}}^{\rm{M}}}}} \right)}^2} + {{\left( {{{\xi _{\rm{D}}^i} \mathord{\left/
 {\vphantom {{\xi _{\rm{D}}^i} {\xi _{\rm{D}}^{\rm{M}}}}} \right.
 \kern-\nulldelimiterspace} {\xi _{\rm{D}}^{\rm{M}}}}} \right)}^2} + {{\left( {{{\xi _{\rm{S}}^i} \mathord{\left/
 {\vphantom {{\xi _{\rm{S}}^i} {\xi _{\rm{S}}^{\rm{M}}}}} \right.
 \kern-\nulldelimiterspace} {\xi _{\rm{S}}^{\rm{M}}}}} \right)}^2}} }}\\
\chi _{\rm{S}}^i = \frac{{{{\xi _{\rm{S}}^i} \mathord{\left/
 {\vphantom {{\xi _{\rm{S}}^i} {\xi _{\rm{S}}^{\rm{M}}}}} \right.
 \kern-\nulldelimiterspace} {\xi _{\rm{S}}^{\rm{M}}}}}}{{\sqrt {{{\left( {{{\xi _{\rm{R}}^i} \mathord{\left/
 {\vphantom {{\xi _{\rm{R}}^i} {\xi _{\rm{R}}^{\rm{M}}}}} \right.
 \kern-\nulldelimiterspace} {\xi _{\rm{R}}^{\rm{M}}}}} \right)}^2} + {{\left( {{{\xi _{\rm{D}}^i} \mathord{\left/
 {\vphantom {{\xi _{\rm{D}}^i} {\xi _{\rm{D}}^{\rm{M}}}}} \right.
 \kern-\nulldelimiterspace} {\xi _{\rm{D}}^{\rm{M}}}}} \right)}^2} + {{\left( {{{\xi _{\rm{S}}^i} \mathord{\left/
 {\vphantom {{\xi _{\rm{S}}^i} {\xi _{\rm{S}}^{\rm{M}}}}} \right.
 \kern-\nulldelimiterspace} {\xi _{\rm{S}}^{\rm{M}}}}} \right)}^2}} }},
\end{aligned}
\end{equation}
where $\xi_{{\rm{R}}}^i$ and $\xi_{\rm{R}}^{\rm{M}}$ denote the value of rotation transformation factor for atom $i$ and the mean value of atoms residing in the matrix, respectively. Thus, we can see that the formula $\xi_{{\rm{R}}}^i/\xi_{\rm{R}}^{\rm{M}}$ indicates the degree of rotation localization for $i$th atom. Similar definitions are used for the other two transformation modes, which are given as $\xi_{{\rm{D}}}^i/\xi_{{\rm{D}}}^{\rm{M}}$ and $\xi_{{\rm{S}}}^i/\xi_{{\rm{S}}}^{\rm{M}}$, respectively. Therefore, we can obtain the relative contribution that each type of event makes to the whole local deformation according to Eq. (\ref{eq:9}). By using such definitions, we then calculate the participation fraction as ${\chi _{\rm{R}}}$, ${\chi _{\rm{D}}}$, and ${\chi _{\rm{S}}}$ for all of the atoms. The maximum value in the population $\{{\chi _{\rm{R}}^i}, {\chi _{\rm{D}}^i}, {\chi _{\rm{S}}^i}\}$ defines an atomic-scale deformation region to be either in RDZ, DDZ or SDZ, respectively. This means that a specific deformation event dominates the defined local region or an atom. For example, if $\chi_{\rm{R}}^i$ is a maximum in the three parameters, the $i$th atom then belongs to RDZ. Figure \ref{Fig:3}d-e shows the good correspondence between the labels of RDZ, DDZ as well as SDZ, and the exact locations of rotation, dilatation, and shear events.

We further investigate the structural origin of RDZ, DDZ and SDZ by using the average degree of the five-fold symmetry (L5FS) parameter \cite{Peng2011,Spaepen2002,Hu2015}. Fig. S3a \cite{sm} shows the average magnitudes of the local L5FS in the RDZ, DDZ and SDZ, respectively, at different levels of deformation. It shows that the highest L5FS level appears in RDZ, compared with those in DDZ and SDZ. This indicates that rotation events are best correlated with the solid-like regions, which are generally more resistant to deformation. This observation is further verified by the evolution of the effective strain ${\tilde \Lambda }$ residing in RDZ, DDZ and SDZ, as shown in Fig. S3b \cite{sm}. It is seen that atoms in SDZs and DDZs are prone to undergo plastic deformation while atoms in RDZs always experience the lowest level of local strain, and thus can resist more distortions. To elaborate more on the new concepts of SDZ, DDZ and RDZ, we provide statistical analysis on the size distribution of them, respectively. By using Eq. (\ref{eq:9}), each atom is labeled as SDZ/DDZ/RDZ. For a local region with cutoff radius $r$, the ratio $\alpha \left( r \right) = (\rm{number}\; \rm{of}\; \rm{RDZ}\; \rm{atoms})/(\rm{number}\; \rm{of}\; \rm{total}\; \rm{atoms})$ is calculated. The involved atoms in a RDZ can thus be captured by gradually increasing the cutoff radius. Once $\alpha (r) < 0.8$, the cutoff and the participated atoms are recorded. Here, the threshold value 0.8 is used to make sure that rotation indeed dominates the exact local region. Similar definitions are used to measure the cluster size of DDZ and SDZ. The results are shown in Fig. S4 \cite{sm}, where the cumulative frequency as well as frequency are plotted as a function of the number of participating atoms and cluster volume. All the size distributions exhibit an exponential-decay-like distribution, with a large fraction of the clusters of tens of atoms. According to the result shown by the cumulative frequency, however, there is somewhat local events containing several hundred atoms. This observation is consistent with the widely accepted size analysis about STZ which contains atoms between a few to approximately several hundreds \cite{Argon1979,Zink2006,Hufnagel2016,Ruan2022}. This indicates that SDZ, DDZ and RDZ are comparable in size to the well-known concept of STZ.

\section{Results and discussions}

\subsection{Atomistic scale pattern of shear banding}
\begin{figure*}
  \centering
  \includegraphics[width=1.0\textwidth]{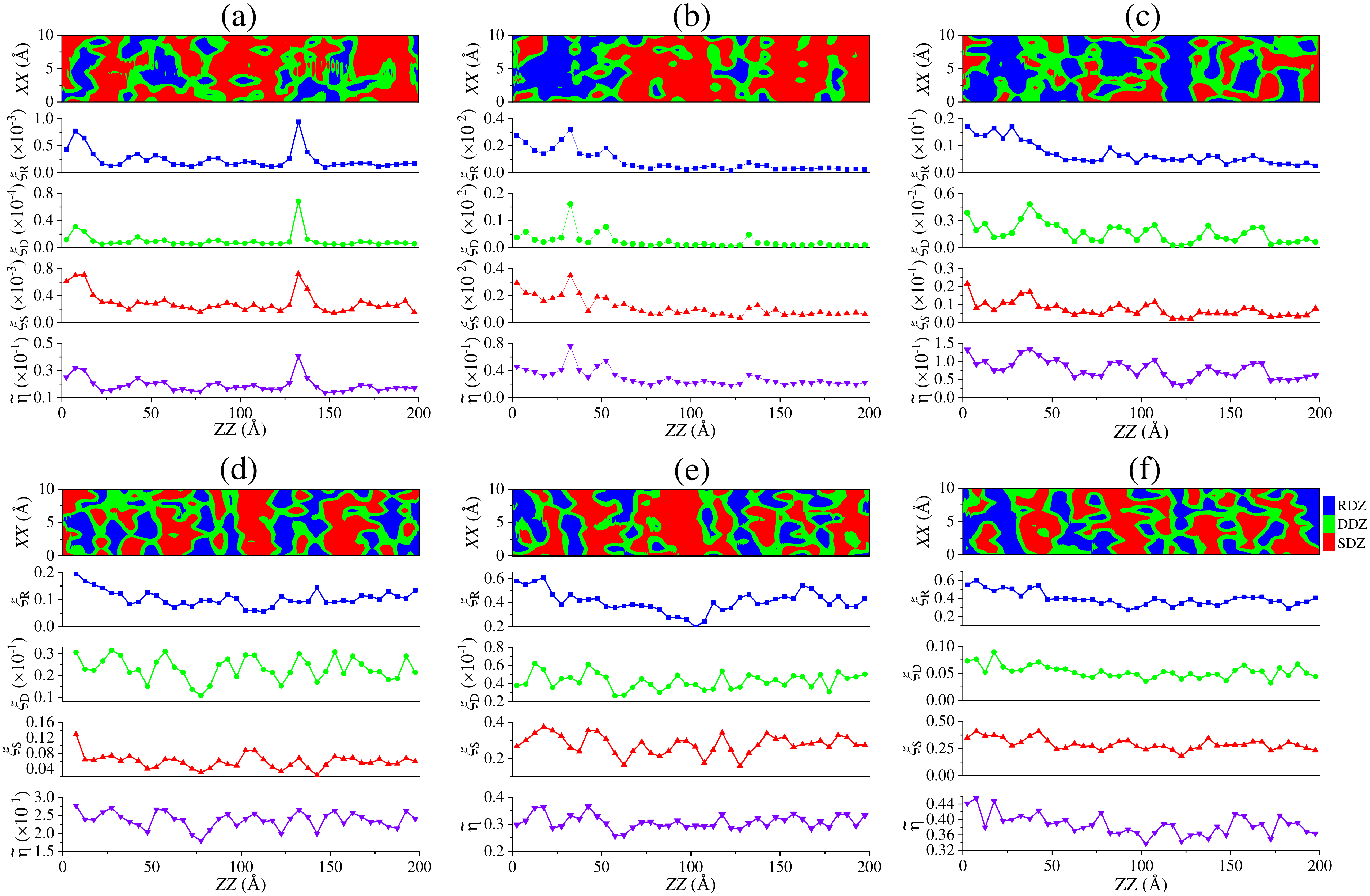}
  \caption{Spatial and temporal evolution of RDZ, DDZ, and SDZ, and the corresponding fluctuation of ${\xi _{\rm{R}}}$, ${\xi _{\rm{D}}}$, ${\xi _{\rm{S}}}$ and ${\tilde \eta }$ along the shear band direction. The snapshots are shown at different strain magnitudes of (a) 0.02, (b) 0.04, (c) 0.06. (d) 0.07, (e) 0.08, and (f) 0.10. Mature shear band forms at strain of 0.08, as demonstrated in Fig. 1a. A slice of 5 {\AA} along the y direction is coarse-grained for the plot.}\label{Fig:4}
\end{figure*}
Having identified the basic deformation units as RDZ, DDZ, and SDZ, it is now possible to discuss the formation process of shear band from the perspective of the spatiotemporal evolution of these events. A close inspection of the temporal evolution of RDZs, DDZs and SDZs maps during shear banding are clarified in Figure \ref{Fig:4}. Here, we can observe that at the strain level of 0.02, 0.04, and 0.06 (Fig. \ref{Fig:4}a-c), RDZs, DDZs, and SDZs are distributed homogeneously and diffusely. This pattern is similar to the laminar-like flows in fluids. It indicates the synchronous activation and motion of rotation, dilatation, and shear events at the early stage of deformation before the appearance of strain localization. However, when the strain reaches 0.07 (Fig. \ref{Fig:3}d), at which point shear band is about to emerge, the laminar-like motion is broken down with the RDZs and SDZ-DDZs distributed separately, which is in striking resemblance with the so-called turbulent flow in flow dynamics \cite{Swinney1978,Falkovich2006}. It implies the consistence between shear banding emergence in metallic glasses with laminar-turbulent transition in liquid, with the same inhomogeneous source -- disordered structure. As the deformation proceeds, the shear band forms in the direction of the maximum shear stress with again the homogenous distribution of RDZs, DDZs, and SDZs as the patterns shown in Fig. \ref{Fig:4}e-f. To obtain further quantitative insight into the temporal and spatial evolution of the deformation pattern, we bin and reduce the values of $\xi_{\rm{R}}$, $\xi_{\rm{D}}$, and $\xi_{\rm{S}}$ along the shear band direction at different strain levels, i.e., 0.02 and 0.04 (early deformation stage), 0.06 (just before strain localization), 0.07 (strain localization), 0.08 (mature shear band formation) and 0.1 (steady-state flow), as shown in Fig. \ref{Fig:4}, respectively, below the deformation pattern. At the early stage of deformation, these three deformation events are homogeneously distributed, and shear, dilatation as well as rotation events overlap with each other in the whole sample. However, one can observe alternative correspondence between the peaks of $\xi_{\rm{R}}$ and valleys of $\xi_{\rm{D}}$ and $\xi_{\rm{S}}$, at strain of 0.07 at which point strain localization has already formed. All of these evidences verify the existence of transition from laminar-like motion to turbulent-like motion which occurs before the onset of shear banding. In addition, the fluctuation of effective strain gradient ${\tilde \eta }$ is also plotted in Fig. \ref{Fig:4} to detail the evolution of strain gradient distribution. It shows that the fluctuation of strain gradient field overlaps with that of $\xi_{\rm{D}}$ and $\xi_{\rm{S}}$ which implies the accumulation of plastic strain in DDZs and SDZs.

\begin{figure}
  \centering
  \includegraphics[width=0.5\textwidth]{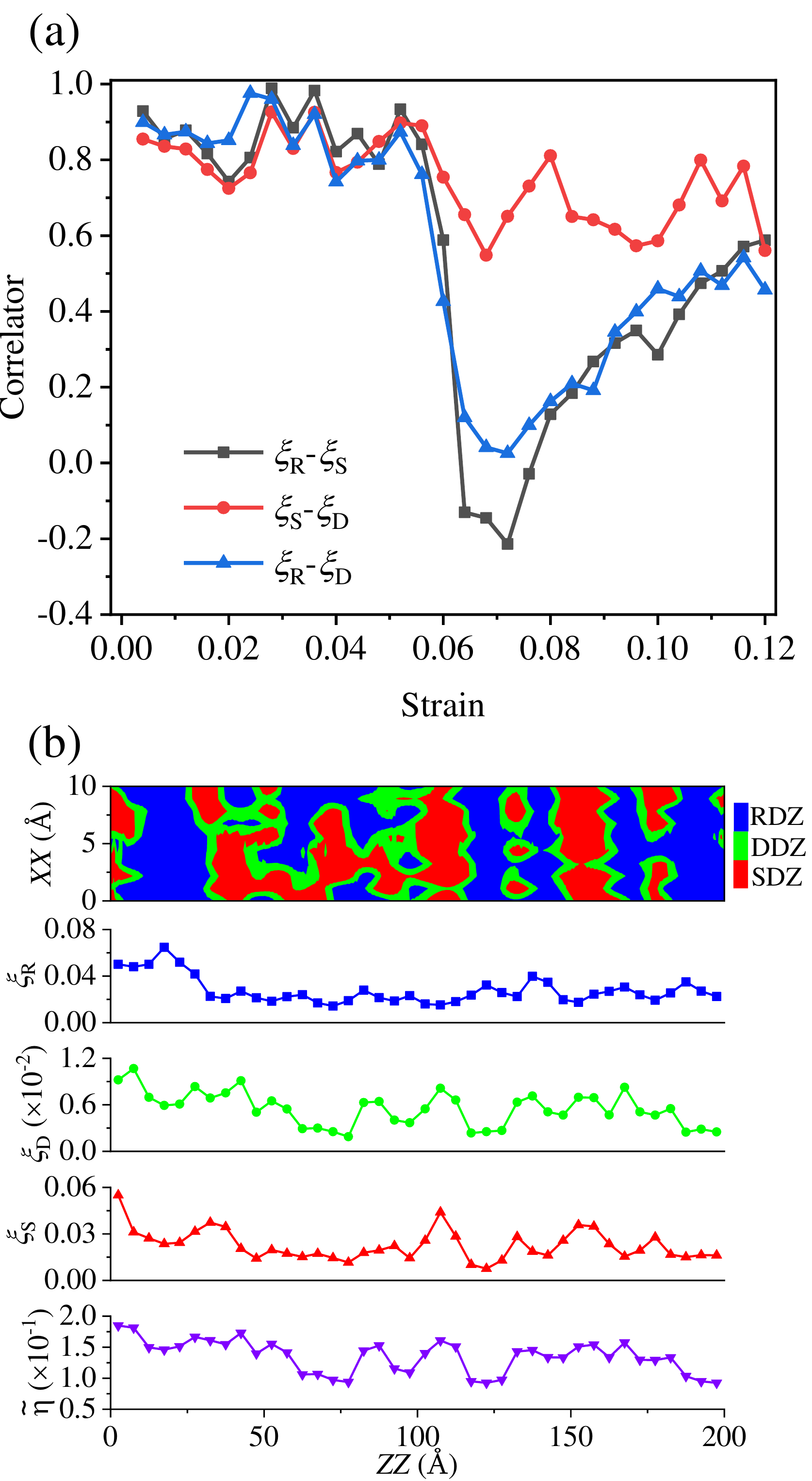}
  \caption{(a) The evolution of correlation among shear, dilatation and rotation during deformation. All of the data used for calculation is taken from shear band. (b) Spatial distribution of RDZ, DDZ, and SDZ, and their fluctuation along the shear band direction at the strain of 0.064.}\label{Fig:5}
\end{figure}

Then, it is necessary to figure out when does the transition from laminar-like motion to turbulent-like motion of shear, dilatation and rotation events happens and what is the physical origin hidden in such atomic motion. To settle such problem, the correlation coefficient is applied to evaluate the evolution of correlation among $\xi_{\rm{R}}$, $\xi_{\rm{D}}$ and $\xi_{\rm{S}}$ in shear band as $C = \frac{{\left\langle {{P_1}{P_2}} \right\rangle  - \left\langle {{P_1}} \right\rangle \left\langle {{P_2}} \right\rangle }}{{\sqrt {\left\langle {P_1^2} \right\rangle  - {{\left\langle {{P_1}} \right\rangle }^2}} \sqrt {\left\langle {P_2^2} \right\rangle  - {{\left\langle {{P_2}} \right\rangle }^2}} }}$. Here $P_1$ and $P_2$ denote two different parameters which can be replaced by $\xi_{\rm{R}}$, $\xi_{\rm{D}}$ and $\xi_{\rm{S}}$. The calculated correlation functions are shown in Fig. \ref{Fig:5}a as a function of macroscopic strain. The positive magnitude of the correlators for $\xi_{\rm{R}}-\xi_{\rm{S}}$ and $\xi_{\rm{R}}-\xi_{\rm{D}}$ during the early stage of deformation indicates the strong positive correlation between shear-dilatation and rotation. This is consistent with the phenomena that these three atomic events overlap with each other during the early stage of deformation. The abrupt decrease of the correlator of $\xi_{\rm{R}}-\xi_{\rm{S}}$ and $\xi_{\rm{R}}-\xi_{\rm{D}}$ from strain = 0.06 to strain = 0.064 shown in Fig. \ref{Fig:5}a clarifies the laminar-turbulent like transition with alternating distribution sequence between shear-dilatation and rotation. Thus, it can be concluded here that the separating distribution of shear-dilatation and rotation occurs at strain of 0.064. To confirm this, the spatial distribution of RDZs, DDZs, and SDZs and the corresponding fluctuations of $\xi_{\rm{R}}$, $\xi_{\rm{D}}$, $\xi_{\rm{S}}$ and ${\tilde \eta }$ along the direction of shear band at strain of 0.064 are plotted in Fig. \ref{Fig:5}b.

\begin{figure}
  \centering
  \includegraphics[width=0.5\textwidth]{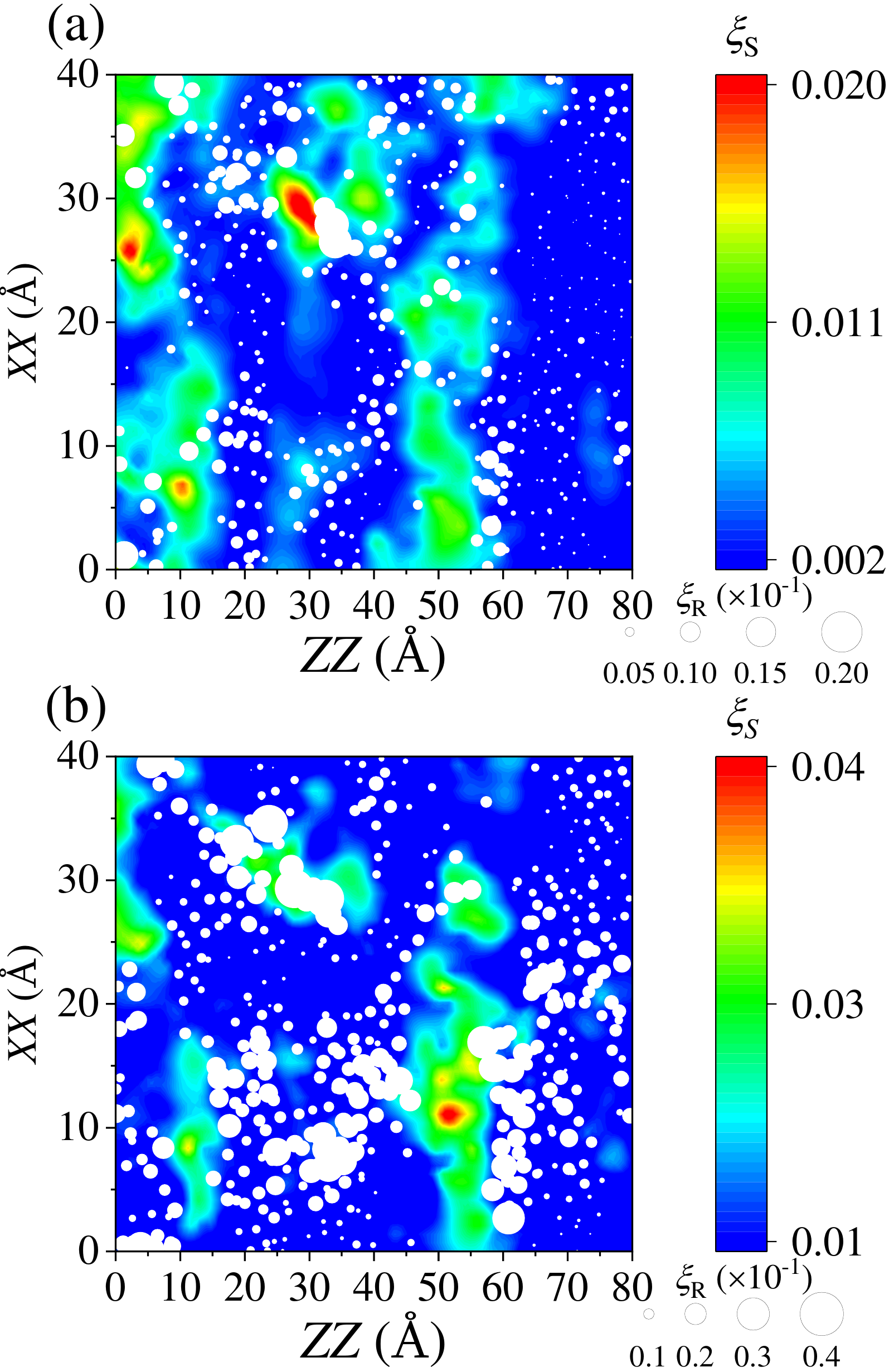}
  \caption{Rotation degree of RDZ between two shear-dilatation regions are enhanced during short period strain = 0.06 to strain = 0.064. The contour maps show the distribution of ${\xi _{\rm{S}}}$ at strain of 0.06 (a) and strain of 0.064 (b) respectively. The white spheres superimposed on the contour maps indicate the positions of the atoms that are labeled RDZs. The size of white spheres denotes magnitude of ${\xi _{\rm{R}}}$.}\label{Fig:6}
\end{figure}

The next task at hand is to figure out what causes the transition from laminar-like to turbulent-like motion during such short period. The colored maps of $\xi_{\rm{S}}$ for atoms inside shear band at strain of 0.06 and 0.064 are plotted in Fig. \ref{Fig:6}a and Fig. \ref{Fig:6}b, respectively. For a direct comparison, the instant RDZs are superimposed in Fig. \ref{Fig:6}a-b, where white spheres, with its size proportional to the magnitude of $\xi_{\rm{R}}$, represent the atoms that have been labeled RDZs. It shows that the distribution of $\xi_{\rm{S}}$ is almost the same between snapshot of 0.06 and 0.064, while the rotation degree for RDZs adjacent shear-dilatation zones is strongly enhanced during this short period, thus resulting in the alternative distribution of shear-dilatation and rotation at the strain of 0.064. Actually, the process of enhancing rotation have taken place earlier, probably at strain of 0.05 when the $\xi_{\rm{R}}-\xi_{\rm{D}}$ and $\xi_{\rm{R}}-\xi_{\rm{S}}$ correlators begin to attenuate. It is at the short period of 0.06 - 0.064 that rotation increasing behavior becomes obvious. Since RDZs are strongly correlated with initial hard clusters, it indicates that solid-like regions along the shear band direction would undergo second activation of rotation, which is more severe than the first one, before softening. Such enhancement of rotation event is similar to the vortex stretching, which usually contributes to the energy dissipation process in turbulent flows \cite{Ooi1999}. In this sense, this class of severe rotation plays an important role in dissipating energy and further softening in the exact regions. A similar pattern is further given by Fig. S5 \cite{sm}, in which spatial distribution of $\xi_{\rm{R}}$, $\xi_{\rm{D}}$, and $\xi_{\rm{S}}$ at strain of 0.06 and 0.064 are plotted. It shows that rotation degree of regions surrounding shear and dilatation zones are enhanced, while the distribution of $\xi_{\rm{S}}$ is almost as it is at strain of 0.06. It is more interesting to find the somewhat enhancement of dilatation residing in the newly activated RDZs. This is the direct evidence of dilatation softening of RDZs which is followed by a percolating process and thus forming shear band. Such percolating mechanism will be discussed in detail in the following section.

\subsection{Extreme statistics of initial plastic events}

\begin{figure*}
  \centering
  \includegraphics[width=0.8\textwidth]{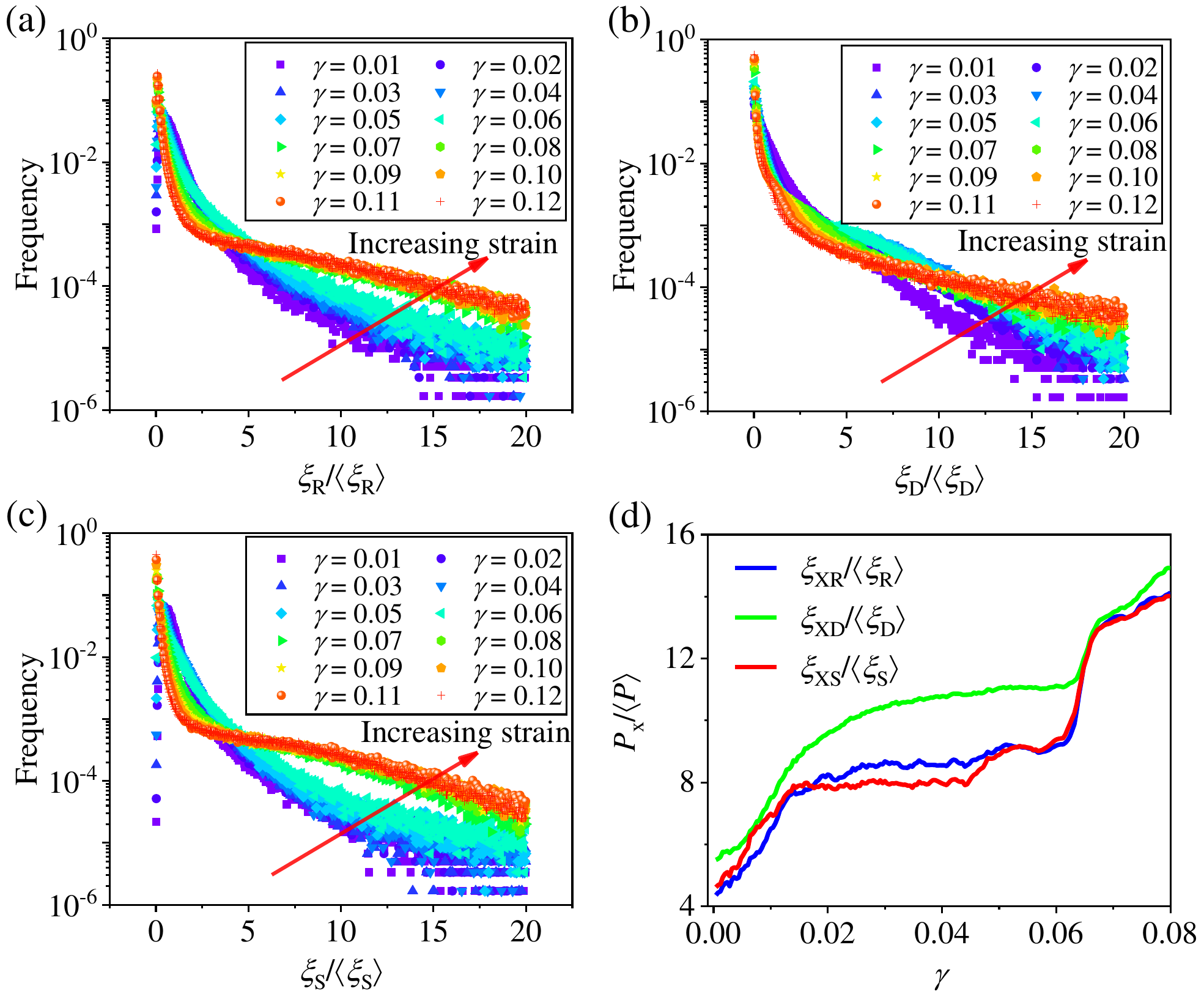}
  \caption{Extreme value analysis of rotation, dilatation and shear events in metallic glass. (a)-(c) Statistical distribution of the extreme values normalized by the mean value for rotation (a), dilatation (b), and shear (c) at various applied strains. (d) The evolution of the normalized extreme value, ${{{\xi _{{\rm{XR}}}}}/{ \left\langle {{\xi _{{\rm{R}}}}} \right\rangle }}$, ${{{\xi _{{\rm{XD}}}}}/{ \left\langle {{\xi _{{\rm{D}}}}} \right\rangle }}$, and ${{{\xi _{{\rm{XS}}}}}/{ \left\langle {{\xi _{{\rm{S}}}}} \right\rangle }}$ during shear banding process.}\label{Fig:7}
\end{figure*}

As mentioned above, the picture is clear that the secondary activation of rotation events (here, we call the rotation together with shear, dilatation at the early stage as the first one) and the following softening of RDZs at the onset of shear banding. However, the precious identification of initial activation of distorting units via shear, dilatation and rotation at earlier stage still remains elusive.

To clarify the respective roles of shear, dilatation and rotation in this process, we apply the extreme value theory (EVT) \cite{1988Extreme,Zhao2013,Cao2019} to analyze the localization behavior of these three deformation events and how they varies with macroscopic strain. First of all, the transformation factors $\xi_{\rm{R}}$, $\xi_{\rm{D}}$ and $\xi_{\rm{S}}$ are rescaled by their mean value $\langle{\xi_{\rm{R}}}\rangle$, $\langle{\xi_{\rm{R}}}\rangle$ and $\langle{\xi_{\rm{R}}}\rangle$, respectively. Fig. \ref{Fig:7}a-c shows the probability density distribution of these reduced factors at various applied strains. It shows that the peak probability always stays at $\sim$ 1 without obvious transitions even for the maximum applied macroscopic strain. The main difference is the much longer tail of distribution as the strain increase. Such long tails at extreme sites for all of distributions of $\xi_{\rm{R}}/\langle{\xi_{\rm{R}}}\rangle$, $\xi_{\rm{D}}/\langle{\xi_{\rm{D}}}\rangle$ and $\xi_{\rm{S}}/\langle{\xi_{\rm{S}}}\rangle$ indicates the onset of inhomogeneous flow for rotation, dilatation and shear units. It is noteworthy that this long-tailed distribution is reminiscent to the non-Gaussianity of velocity gradient distributions observed in turbulence via high-resolution direct numerical simulation \cite{Gotoh2002,Ishihara2007,Ishihara2009}. It reinforces that the emergence of shear banding in metallic glasses is similar to the transition from homogeneous laminar motion to inhomogeneous turbulent flow in the community of fluid dynamics.

To further quantify the evolution of such long-tailing behavior for rotation, dilatation and shear respectively, we track the extreme sites residing in the long tail. Atoms with the highest 1\% value of transformation factors are picked out. The reduced mean value of these atoms ($P_{\rm{X}}/\langle{P}\rangle$, where $P_{\rm{X}}$ denotes the extreme values of $P$ and $P$ can be replaced by $\xi_{\rm{R}}$ for rotation, $\xi_{\rm{D}}$ for dilatation and $\xi_{\rm{S}}$ for shear) are used to measure the degree of heterogeneity for different deformation events. The evolution of $\xi_{\rm{XR}}/\langle{\xi_{\rm{R}}}\rangle$, $\xi_{\rm{XD}}/\langle{\xi_{\rm{D}}}\rangle$ and $\xi_{\rm{XS}}/\langle{\xi_{\rm{S}}}\rangle$ as functions of macroscopic strain are shown in Fig. \ref{Fig:7}d. It shows that the long-tail behavior for all of rotation, dilatation and rotation events experience two step growth before formation of mature shear band at strain of 0.08. One is at the embryonic stage of deformation characterizing the initial activation of rotation, dilatation and shear. The other takes place at the critical time of percolation which will be discussed in detail in the sext subsection. The long plateaus adjacent these two abrupt enhancement implies the cage effect or percolation barrier before shear banding. As $P_{\rm{X}}/\langle{P}\rangle$ is dimensionless, we can compare the degree of rotation, dilatation and shear contributing to the whole localization behavior directly. It is interesting to find out the most important role of dilatation over rotation and shear at the early stage of deformation. This is the direct evidence to verify the dominating role played by dilatation during early deformation. Based on this evidence, the scenario for initial deformation process emerges: when applied strain, soft regions in metallic glass mainly undergo the dilatation event and generate free volume, within which atoms are permitted to do further rotation and shear motion, thus causing the activation of initial plastic units. To test whether the stress concentration introduced by notch would influence the operations of SDZ, DDZ and RDZs, we conduct an additional tensile test on a glass sample without notch. The simulation results are shown in Fig. S6 \cite{sm}. It shows that the magnitude of stress of the glass without notch is much higher than that of the notched sample. Besides, the yielding behavior of sample without notch takes place later than the notched one. According to the extreme value analysis shown in Fig. S6b \cite{sm}, dilatation localization also dominates the initial plastic events in the glass sample without notch. It indicates that the notch will not affect the initial operations of shear, dilatation, and rotation. It is an intrinsic nature of the plastic deformation in metallic glasses that dilatation plays a dominating role at the early stage of deformation. However, the stress concentration induced by the notch will influence the nucleation site of the initial activation of dilatation, as evidenced by Fig. S7 \cite{sm} where spatial distribution of the dilatation transformation factor as well as mean stress are plotted together. It shows that the nucleation of the initial dilatation events takes place near the notch where stress concentration occurs.

\subsection{Percolation transition and shear banding emergence}

\begin{figure*}
  \centering
  \includegraphics[width=1.0\textwidth]{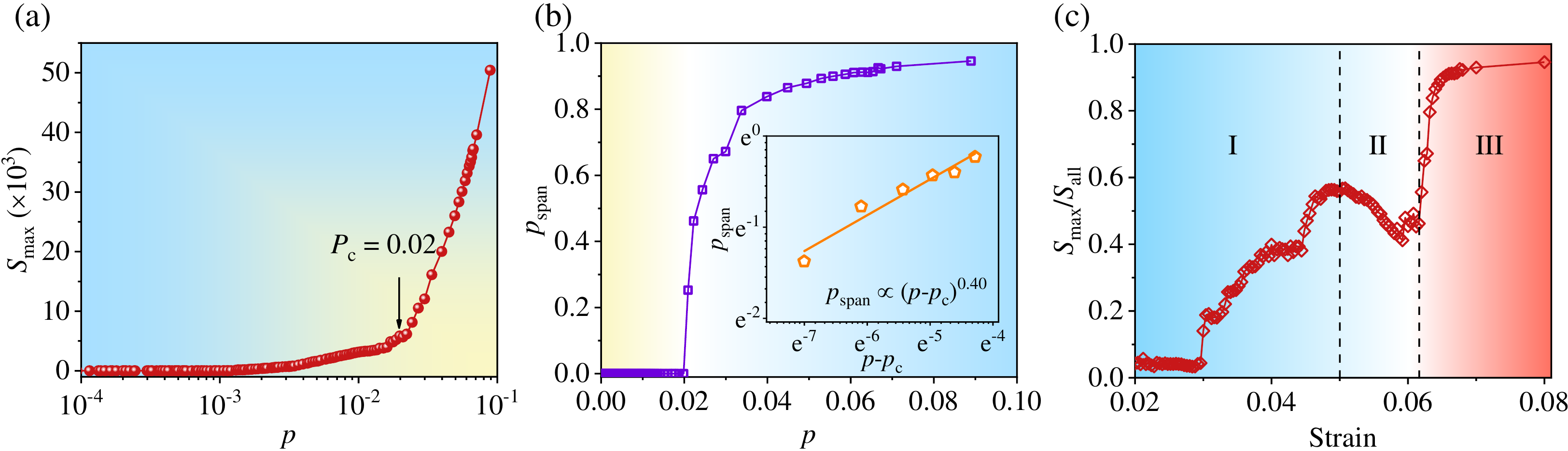}
  \caption{Percolation analysis in metallic glass. (a) Size of the maximum percolating cluster as a function of $p$. Here, the critical fraction $p_c$ is characterized as 0.02. (b) Variation of percolating probability, $p_{\rm{span}}$, with $p$. The inset shows a power-law statistic, ${p_{{\rm{span}}}} \propto {\left( {p - {p_c}} \right)^\beta }$ with exponent 0.40 which is consistent with the classical scaling behavior in three dimensional percolation analysis \cite{Stauffer1994}. (c) Time evolution of percolating factor, $S_{\rm{max}}/S_{\rm{all}}$, displaying three characteristic flow steps before shear banding formation.}\label{Fig:8}
\end{figure*}
The remaining question is how these co-existing localized plastic regions, dominated by the initial dilatation events, span to large scales in the form of shear bands. To settle this problem, we apply percolation analysis which usually provides a complementary framework for analysing shear banding nature in metallic glass and other disordered materials \cite{Shrivastav2016,Cao2018,Cao2019}. Firstly, we define an atom to be plastic or not, by setting a threshold 0.25 of effective atomic strain ${\tilde \Lambda }$. The snapshots of plastic atoms at various applied strains are shown in Fig. S8 \cite{sm}, where one can observe a clear transition from stochastic activation to percolation of plastic regions with increasing strain. To quantify and characterize the occurrence of percolation transition, we monitor the fraction of plastic atoms, $p$, as well as the number of plastic atoms residing in the maximum percolating cluster, $S_{\rm{max}}$. Fig. \ref{Fig:8}a plots $S_{\rm{max}}$ as a function of $p$, which shows that beyond a critical fraction $p_c = 0.02$ tremendous amounts of plastic atoms are covered in the maximum cluster. This indicates the occurrence of the percolation transition. It should be noted here that $p_c = 0.02$ is smaller than the result of classical percolation analysis for three dimension \cite{Stauffer1994}. This is due to the notched sample used in the present work that leads to stress concentration near the notched region, and thus bringing the transition forward. Above the percolation threshold, we can figure out the spanning cluster which is exactly the shear band. Here, we compute the probability that a plastic atom belongs to the spanning cluster, namely $p_{\rm{span}}$. Then we plot in Fig. \ref{Fig:8}b the evolution of $p_{\rm{span}}$ as a function of $p$, which shows that $p_{\rm{span}}$ will dramatically increase with exponential dependence on $p$ after the critical point $p_c = 0.02$. Then, it quickly converges to $\sim$ 1, indicating that all the plastic atoms are part of such a spanning cluster or shear band. In the inset of Fig. \ref{Fig:8}b, we plot the power-law statistics when $p$ approaches $p_c$ from above. Here, it is evident to observe the power-law scaling behavior ${p_{{\rm{span}}}} \propto {\left( {p - {p_c}} \right)^\beta }$ with the critical exponent of 0.4, which is consistent with the classical scaling nature in three dimensional percolating analysis \cite{Stauffer1994} as it is a universal value only depending on the dimension of sample. This is the direct evidence that the connection of existing localized plastic units, manifesting as shear banding emergence, follows the mechanism of classical percolation transition.

To characterize the accurate physical picture of shear banding emergence, we further investigate the time evolution of percolating process evaluated by the fraction of largest percolating cluster in all plastic atoms, $S_{\rm{max}}/S_{\rm{all}}$, namely percolating factor. Here, $S_{\rm{all}}$ denotes the number of atoms that have undergone plastic events at instant time, while $S_{\rm{max}}$ characterizes the size of percolating cluster. Thus the abrupt increase of $S_{\rm{max}}/S_{\rm{all}}$ can catch the occurrence of percolating events. Fig. \ref{Fig:8}c plots the time evolution of percolating factor $S_{\rm{max}}/S_{\rm{all}}$, which shows three characteristic regimes before shear banding formation. In regime I, dilatation drives the stochastic activation of liquid-like regions as mentioned above, thus $S_{\rm{max}}/S_{\rm{all}}$ increases as strain proceeds, with two step growth characterizing the local percolating events near the notch region. In regime II, $S_{\rm{max}}/S_{\rm{all}}$ decays with the increasing strain. It indicates that the percolating process in shear band is sluggish while newly activation events in the matrix still goes on, causing the growth speed of percolating cluster less than that of stochastic activation. The hidden mechanism is that the solid-like clusters near the co-existing activated regions function as obstacles hindering the broadening of plastic events. As aforementioned, these solid-like clusters adjacent initial plastic regions will be characterized as the secondary activated RDZs with the rotation degree dramatically enhanced during this period, as shown in Fig. \ref{Fig:5}. This in turn enhances the softening of hard RDZs, which is followed by the percolating transition at the critical point as shown in the regime III. Finally $S_{\rm{max}}/S_{\rm{all}}$ converges to $\sim$ 1, implying the formation of mature shear band.

\subsection{Comprehensive mechanism of shear band}

Having identified the temporal and spatial sequence of elementary units, dilatation, rotation and shear in shear banding, it is now possible to discuss the comprehensive mechanism of shear band formation led by collective motion of shear, dilatation and rotation events with the help of EVT and percolation analysis mentioned above. In this connection, the shear banding process can be separated into four critical steps. In step 1, owing to the disordered structure of metallic glasses, the atoms can exhibit a variety of local environments, causing the inhomogeneous distribution of stable and unstable clusters. Then, as the strain proceeds, dilatation events of soft regions are activated, causing the dilatation localization over the matrix, which is recognized as step 2. The dominated role played by dilatation during the initial activation of plastic events have been verified by the extreme value analysis, see Fig. \ref{Fig:7}. Within the enough free volume induced by dilatation events, atoms residing in the dilating regions are able to undergo further shear and rotation motion. In this period, the spatial distribution of dilatation, shear and rotation are nesting with each other, see Fig. \ref{Fig:4}a-c. Such distorting units containing dilatation, shear and rotation are actually the familiar concept of STZ. However, the decoupling of entangled three events makes the physical process of stochastic activation more clear, where the picture of dilatation pioneering shear and rotation is firstly proposed.

In step 3, the further broadening of plastic deformation in the soft regions is hindered by the surrounding hard clusters, causing the accumulation of strain gradient in the initial plastic regions. As evidenced by the structural feature of RDZs in Fig. S3 \cite{sm}, atoms in hard clusters mainly undertake rotating motion. With strain goes on, rotation degree of those solid-like clusters adjacent initial plastic regions is dramatically enhanced, as shown in Fig. \ref{Fig:6}. In contrast to the initial activation of rotation events together with shear and dilatation in soft regions, this secondary activation of rotation is stronger which constructs the concept of RDZ in the present work. Therefore, we can visualize the correlation decrease of $\xi_{\rm{R}}-\xi_{\rm{D}}$ and $\xi_{\rm{R}}-\xi_{\rm{S}}$, and as a consequence, the transition from laminar-like motion to turbulent-like motion of SDZ-DDZs and RDZs during such short time period, see Fig. \ref{Fig:4}d and Fig. \ref{Fig:5}. Such rotation motion, much like the vortex in turbulent flow \cite{Ooi1999,Swinney1978,Falkovich2006}, in turn contributes to the energy cascade process and further aggravates the dilatation of the border of these RDZs. This is verified by Fig. \ref{Fig:4} where the boundary of RDZs is entangled with DDZs. It figures out the critical role of rotation during the further softening of hard clusters along the direction of shear banding. In the final step 4, when the softening of RDZs reaches the limit, that is at the critical point of percolating, the well-activated distorting units will connect with each other, driving the emergence of shear band in the direction of maximum stress.

\section{Conclusion}
In summary, we propose a theoretical TTG framework by categorizing the displacement gradient tensor and strain gradient tensor into three parts, and thus decouple the entangled shear, dilatation, and rotation events hidden in the deformation of MGs. The proposed TTG model in this work simultaneously capture the mechanism from both affine and nonaffine perspectives. This unprecedented combination gives rise to a more comprehensive and more effective description of the cooperative shear, dilatation and rotation events beyond the conventional, purely affine or non-affine model. The proposed transformation factors, $\xi_{\rm{S}}$, $\xi_{\rm{D}}$ and $\xi_{\rm{R}}$ can accurately quantify the degree of shear, dilatation, and rotation events, respectively. From the perspective of the participation fraction, SDZs, DDZs, and RDZs are defined to reveal the physical mechanism of shear band formation. At low strain, SDZs, DDZs and RDZs are strongly coupled with each other, leading to stochastic activation of initial plastic events in soft regions. The EVT results suggest the predominant process during the activation of initial plastic units is the dilatation of liquid-like spots. It is in good agreement with early theoretical works based on continuum mechanics \cite{Greer2013,Jiang2009,Spaepen1977}. The present work further gives the direct atomic evidence that dilatation softening is the root source for the onset of strain localization. It is also found that before the softening of the exact solid-like clusters adjacent initial plastic regions, the rotation degree in these hard regions will be significantly enhanced, causing the secondary activation of RDZs there. This activation of RDZs is at a dominating level compared to the first one and causes the transition from homogeneous laminar-like motion to inhomogeneous turbulent-like flow which is effectively characterized by the separately distributed SDZs, DDZs, and RDZs. This phenomena provides a natural connection between shear banding emergence and liquid flow. It opens up plenty of opportunities for the community to understand the microscopic mechanism of deformation in disordered materials with the help of fluid dynamics theory. Such severely rotating motion further aggravates the softening of hard regions and contributes to the directional perturbation process as well as ultimate shear band formation. Our percolation analysis yields the critical power-law scaling nature of shear banding, with scaling exponent in analogy with classical percolating theory. It demonstrates how the co-existing localized plastic regions span to large scales in the form of shear banding and confirms the above three-unit atomistic description of shear banding mechanism.

Therefore, we believe that our atomic-scale scenario of shear band emergence may provide a fundamental understanding of the nature of rheological behavior in metallic glass. While this study concentrates on the metallic glass, the TTG model, relying on atomic position alone, can be directly applied to other disordered materials. The incorporation of both affine and non-affine components of deformation may open up new opportunities to gain deeply conceptual understanding not achieved via conventional models.

\begin{acknowledgments}
This work was supported by the NSFC (No. 11790292, No. 12072344), the NSFC Basic Science Center Program for “Multiscale Problems in Nonlinear Mechanics” (No.11988102), National Key Research and Development Program of China (No. 2017YFB0702003). The work was also supported by the Strategic Priority Research Program (No. XDB22040302, XDB22040303), the Key Research Program of Frontier Sciences (No. QYZDJSSWJSC011), and the Youth Innovation Promotion Association of CAS (No. 2017025).
\end{acknowledgments}


\begin{thebibliography}{84}%
\makeatletter
\providecommand \@ifxundefined [1]{%
 \@ifx{#1\undefined}
}%
\providecommand \@ifnum [1]{%
 \ifnum #1\expandafter \@firstoftwo
 \else \expandafter \@secondoftwo
 \fi
}%
\providecommand \@ifx [1]{%
 \ifx #1\expandafter \@firstoftwo
 \else \expandafter \@secondoftwo
 \fi
}%
\providecommand \natexlab [1]{#1}%
\providecommand \enquote  [1]{``#1''}%
\providecommand \bibnamefont  [1]{#1}%
\providecommand \bibfnamefont [1]{#1}%
\providecommand \citenamefont [1]{#1}%
\providecommand \href@noop [0]{\@secondoftwo}%
\providecommand \href [0]{\begingroup \@sanitize@url \@href}%
\providecommand \@href[1]{\@@startlink{#1}\@@href}%
\providecommand \@@href[1]{\endgroup#1\@@endlink}%
\providecommand \@sanitize@url [0]{\catcode `\\12\catcode `\$12\catcode
  `\&12\catcode `\#12\catcode `\^12\catcode `\_12\catcode `\%12\relax}%
\providecommand \@@startlink[1]{}%
\providecommand \@@endlink[0]{}%
\providecommand \url  [0]{\begingroup\@sanitize@url \@url }%
\providecommand \@url [1]{\endgroup\@href {#1}{\urlprefix }}%
\providecommand \urlprefix  [0]{URL }%
\providecommand \Eprint [0]{\href }%
\providecommand \doibase [0]{https://doi.org/}%
\providecommand \selectlanguage [0]{\@gobble}%
\providecommand \bibinfo  [0]{\@secondoftwo}%
\providecommand \bibfield  [0]{\@secondoftwo}%
\providecommand \translation [1]{[#1]}%
\providecommand \BibitemOpen [0]{}%
\providecommand \bibitemStop [0]{}%
\providecommand \bibitemNoStop [0]{.\EOS\space}%
\providecommand \EOS [0]{\spacefactor3000\relax}%
\providecommand \BibitemShut  [1]{\csname bibitem#1\endcsname}%
\let\auto@bib@innerbib\@empty
\bibitem [{\citenamefont {Greer}\ \emph {et~al.}(2013)\citenamefont {Greer},
  \citenamefont {Cheng},\ and\ \citenamefont {Ma}}]{Greer2013}%
  \BibitemOpen
  \bibfield  {author} {\bibinfo {author} {\bibfnamefont {A.}~\bibnamefont
  {Greer}}, \bibinfo {author} {\bibfnamefont {Y.}~\bibnamefont {Cheng}},\ and\
  \bibinfo {author} {\bibfnamefont {E.}~\bibnamefont {Ma}},\ }\bibfield
  {title} {\bibinfo {title} {{Shear bands in metallic glasses}},\ }\href
  {https://doi.org/10.1016/j.mser.2013.04.001} {\bibfield  {journal} {\bibinfo
  {journal} {Mater. Sci. Eng. R}\ }\textbf {\bibinfo {volume} {74}},\ \bibinfo
  {pages} {71} (\bibinfo {year} {2013})}\BibitemShut {NoStop}%
\bibitem [{\citenamefont {Schuh}\ \emph {et~al.}(2007)\citenamefont {Schuh},
  \citenamefont {Hufnagel},\ and\ \citenamefont {Ramamurty}}]{Schuh2007}%
  \BibitemOpen
  \bibfield  {author} {\bibinfo {author} {\bibfnamefont {C.~A.}\ \bibnamefont
  {Schuh}}, \bibinfo {author} {\bibfnamefont {T.~C.}\ \bibnamefont
  {Hufnagel}},\ and\ \bibinfo {author} {\bibfnamefont {U.}~\bibnamefont
  {Ramamurty}},\ }\bibfield  {title} {\bibinfo {title} {{Mechanical behavior of
  amorphous alloys}},\ }\href {https://doi.org/10.1016/j.actamat.2007.01.052}
  {\bibfield  {journal} {\bibinfo  {journal} {Acta Mater.}\ }\textbf {\bibinfo
  {volume} {55}},\ \bibinfo {pages} {4067} (\bibinfo {year}
  {2007})}\BibitemShut {NoStop}%
\bibitem [{\citenamefont {Falk}\ and\ \citenamefont {Langer}(2011)}]{Falk2011}%
  \BibitemOpen
  \bibfield  {author} {\bibinfo {author} {\bibfnamefont {M.~L.}\ \bibnamefont
  {Falk}}\ and\ \bibinfo {author} {\bibfnamefont {J.}~\bibnamefont {Langer}},\
  }\bibfield  {title} {\bibinfo {title} {{Deformation and Failure of Amorphous,
  Solidlike Materials}},\ }\href
  {https://doi.org/10.1146/annurev-conmatphys-062910-140452} {\bibfield
  {journal} {\bibinfo  {journal} {Annu. Rev. Condens. Matter Phys.}\ }\textbf
  {\bibinfo {volume} {2}},\ \bibinfo {pages} {353} (\bibinfo {year}
  {2011})}\BibitemShut {NoStop}%
\bibitem [{\citenamefont {Gourlay}\ and\ \citenamefont
  {Dahle}(2007)}]{Gourlay2007}%
  \BibitemOpen
  \bibfield  {author} {\bibinfo {author} {\bibfnamefont {C.~M.}\ \bibnamefont
  {Gourlay}}\ and\ \bibinfo {author} {\bibfnamefont {A.~K.}\ \bibnamefont
  {Dahle}},\ }\bibfield  {title} {\bibinfo {title} {{Dilatant shear bands in
  solidifying metals}},\ }\href {https://doi.org/10.1038/nature05426}
  {\bibfield  {journal} {\bibinfo  {journal} {Nature}\ }\textbf {\bibinfo
  {volume} {445}},\ \bibinfo {pages} {70} (\bibinfo {year} {2007})}\BibitemShut
  {NoStop}%
\bibitem [{\citenamefont {Schall}\ and\ \citenamefont {{Van
  Hecke}}(2010)}]{Schall2010}%
  \BibitemOpen
  \bibfield  {author} {\bibinfo {author} {\bibfnamefont {P.}~\bibnamefont
  {Schall}}\ and\ \bibinfo {author} {\bibfnamefont {M.}~\bibnamefont {{Van
  Hecke}}},\ }\bibfield  {title} {\bibinfo {title} {{Shear bands in matter with
  granularity}},\ }\href {https://doi.org/10.1146/annurev-fluid-121108-145544}
  {\bibfield  {journal} {\bibinfo  {journal} {Annu. Rev. Fluid Mech.}\ }\textbf
  {\bibinfo {volume} {42}},\ \bibinfo {pages} {67} (\bibinfo {year}
  {2010})}\BibitemShut {NoStop}%
\bibitem [{\citenamefont {Furukawa}\ and\ \citenamefont
  {Tanaka}(2009)}]{Furukawa2009}%
  \BibitemOpen
  \bibfield  {author} {\bibinfo {author} {\bibfnamefont {A.}~\bibnamefont
  {Furukawa}}\ and\ \bibinfo {author} {\bibfnamefont {H.}~\bibnamefont
  {Tanaka}},\ }\bibfield  {title} {\bibinfo {title} {{Inhomogeneous flow and
  fracture of glassymaterials}},\ }\href {https://doi.org/10.1038/nmat2468}
  {\bibfield  {journal} {\bibinfo  {journal} {Nat. Mater.}\ }\textbf {\bibinfo
  {volume} {8}},\ \bibinfo {pages} {601} (\bibinfo {year} {2009})}\BibitemShut
  {NoStop}%
\bibitem [{\citenamefont {Bai}\ and\ \citenamefont {Dodd}(1992)}]{Bai1992}%
  \BibitemOpen
  \bibfield  {author} {\bibinfo {author} {\bibfnamefont {Y.}~\bibnamefont
  {Bai}}\ and\ \bibinfo {author} {\bibfnamefont {B.}~\bibnamefont {Dodd}},\
  }\href@noop {} {\emph {\bibinfo {title} {Adiabatic Shear Localization}}}\
  (\bibinfo  {publisher} {Pergamon Press},\ \bibinfo {address} {Oxford},\
  \bibinfo {year} {1992})\BibitemShut {NoStop}%
\bibitem [{\citenamefont {Wisitsorasaka}\ and\ \citenamefont
  {Wolynes}(2017)}]{Wisitsorasaka2017}%
  \BibitemOpen
  \bibfield  {author} {\bibinfo {author} {\bibfnamefont {A.}~\bibnamefont
  {Wisitsorasaka}}\ and\ \bibinfo {author} {\bibfnamefont {P.~G.}\ \bibnamefont
  {Wolynes}},\ }\bibfield  {title} {\bibinfo {title} {{Dynamical theory of
  shear bands in structural glasses}},\ }\href
  {https://doi.org/10.1073/pnas.1620399114} {\bibfield  {journal} {\bibinfo
  {journal} {Proc. Natl. Acad. Sci. U. S. A.}\ }\textbf {\bibinfo {volume}
  {114}},\ \bibinfo {pages} {1287} (\bibinfo {year} {2017})}\BibitemShut
  {NoStop}%
\bibitem [{\citenamefont {Wang}(2019)}]{Wang2019}%
  \BibitemOpen
  \bibfield  {author} {\bibinfo {author} {\bibfnamefont {W.~H.}\ \bibnamefont
  {Wang}},\ }\bibfield  {title} {\bibinfo {title} {{Dynamic relaxations and
  relaxation-property relationships in metallic glasses}},\ }\href
  {https://doi.org/10.1016/j.pmatsci.2019.03.006} {\bibfield  {journal}
  {\bibinfo  {journal} {Prog. Mater. Sci.}\ }\textbf {\bibinfo {volume}
  {106}},\ \bibinfo {pages} {100561} (\bibinfo {year} {2019})}\BibitemShut
  {NoStop}%
\bibitem [{\citenamefont {Yan}\ \emph {et~al.}(2021)\citenamefont {Yan},
  \citenamefont {Li}, \citenamefont {Xu},\ and\ \citenamefont
  {Meyers}}]{Yan2021}%
  \BibitemOpen
  \bibfield  {author} {\bibinfo {author} {\bibfnamefont {N.}~\bibnamefont
  {Yan}}, \bibinfo {author} {\bibfnamefont {Z.}~\bibnamefont {Li}}, \bibinfo
  {author} {\bibfnamefont {Y.}~\bibnamefont {Xu}},\ and\ \bibinfo {author}
  {\bibfnamefont {M.~A.}\ \bibnamefont {Meyers}},\ }\bibfield  {title}
  {\bibinfo {title} {{Shear localization in metallic materials at high strain
  rates}},\ }\href {https://doi.org/10.1016/j.pmatsci.2020.100755} {\bibfield
  {journal} {\bibinfo  {journal} {Prog. Mater. Sci.}\ }\textbf {\bibinfo
  {volume} {119}},\ \bibinfo {pages} {100755} (\bibinfo {year}
  {2021})}\BibitemShut {NoStop}%
\bibitem [{\citenamefont {Spaepen}(1977)}]{Spaepen1977}%
  \BibitemOpen
  \bibfield  {author} {\bibinfo {author} {\bibfnamefont {F.}~\bibnamefont
  {Spaepen}},\ }\bibfield  {title} {\bibinfo {title} {{A microscopic mechanism
  for steady state inhomogeneous flow in metallic glasses}},\ }\href
  {https://doi.org/10.1016/0001-6160(77)90232-2} {\bibfield  {journal}
  {\bibinfo  {journal} {Acta Metall.}\ }\textbf {\bibinfo {volume} {25}},\
  \bibinfo {pages} {407} (\bibinfo {year} {1977})}\BibitemShut {NoStop}%
\bibitem [{\citenamefont {Langer}(2001)}]{Langer2001}%
  \BibitemOpen
  \bibfield  {author} {\bibinfo {author} {\bibfnamefont {J.~S.}\ \bibnamefont
  {Langer}},\ }\bibfield  {title} {\bibinfo {title} {{Microstructural shear
  localization in plastic deformation of amorphous solids}},\ }\href
  {https://doi.org/10.1103/PhysRevE.64.011504} {\bibfield  {journal} {\bibinfo
  {journal} {Phys. Rev. E}\ }\textbf {\bibinfo {volume} {64}},\ \bibinfo
  {pages} {011504} (\bibinfo {year} {2001})}\BibitemShut {NoStop}%
\bibitem [{\citenamefont {Manning}\ \emph {et~al.}(2009)\citenamefont
  {Manning}, \citenamefont {Daub}, \citenamefont {Langer},\ and\ \citenamefont
  {Carlson}}]{Manning2009}%
  \BibitemOpen
  \bibfield  {author} {\bibinfo {author} {\bibfnamefont {M.~L.}\ \bibnamefont
  {Manning}}, \bibinfo {author} {\bibfnamefont {E.~G.}\ \bibnamefont {Daub}},
  \bibinfo {author} {\bibfnamefont {J.~S.}\ \bibnamefont {Langer}},\ and\
  \bibinfo {author} {\bibfnamefont {J.~M.}\ \bibnamefont {Carlson}},\
  }\bibfield  {title} {\bibinfo {title} {{Rate-dependent shear bands in a
  shear-transformation-zone model of amorphous solids}},\ }\href
  {https://doi.org/10.1103/PhysRevE.79.016110} {\bibfield  {journal} {\bibinfo
  {journal} {Phys. Rev. E}\ }\textbf {\bibinfo {volume} {79}},\ \bibinfo
  {pages} {016110} (\bibinfo {year} {2009})}\BibitemShut {NoStop}%
\bibitem [{\citenamefont {Egami}\ \emph {et~al.}(2013)\citenamefont {Egami},
  \citenamefont {Iwashita},\ and\ \citenamefont {Dmowski}}]{Egami2013}%
  \BibitemOpen
  \bibfield  {author} {\bibinfo {author} {\bibfnamefont {T.}~\bibnamefont
  {Egami}}, \bibinfo {author} {\bibfnamefont {T.}~\bibnamefont {Iwashita}},\
  and\ \bibinfo {author} {\bibfnamefont {W.}~\bibnamefont {Dmowski}},\
  }\bibfield  {title} {\bibinfo {title} {{Mechanical properties of metallic
  glasses}},\ }\href {https://doi.org/10.3390/met3010077} {\bibfield  {journal}
  {\bibinfo  {journal} {Metals (Basel).}\ }\textbf {\bibinfo {volume} {3}},\
  \bibinfo {pages} {77} (\bibinfo {year} {2013})}\BibitemShut {NoStop}%
\bibitem [{\citenamefont {Dai}(2012)}]{DAI2012311}%
  \BibitemOpen
  \bibfield  {author} {\bibinfo {author} {\bibfnamefont {L.~H.}\ \bibnamefont
  {Dai}},\ }\bibfield  {title} {\bibinfo {title} {{8 - Shear Banding in Bulk
  Metallic Glasses}},\ }in\ \href@noop {} {\emph {\bibinfo {booktitle}
  {Adiabatic Shear Localization: Frontiers and Advances}}},\ \bibinfo {editor}
  {edited by\ \bibinfo {editor} {\bibfnamefont {B.}~\bibnamefont {Dodd}}\ and\
  \bibinfo {editor} {\bibfnamefont {Y.}~\bibnamefont {Bai}}}\ (\bibinfo
  {publisher} {Elsevier},\ \bibinfo {address} {Oxford},\ \bibinfo {year}
  {2012})\ \bibinfo {edition} {2nd}\ ed.,\ pp.\ \bibinfo {pages}
  {311--361}\BibitemShut {NoStop}%
\bibitem [{\citenamefont {Cheng}\ and\ \citenamefont {Ma}(2011)}]{Cheng2011}%
  \BibitemOpen
  \bibfield  {author} {\bibinfo {author} {\bibfnamefont {Y.~Q.}\ \bibnamefont
  {Cheng}}\ and\ \bibinfo {author} {\bibfnamefont {E.}~\bibnamefont {Ma}},\
  }\bibfield  {title} {\bibinfo {title} {{Atomic-level structure and
  structure-property relationship in metallic glasses}},\ }\href
  {https://doi.org/10.1016/j.pmatsci.2010.12.002} {\bibfield  {journal}
  {\bibinfo  {journal} {Prog. Mater. Sci.}\ }\textbf {\bibinfo {volume} {56}},\
  \bibinfo {pages} {379} (\bibinfo {year} {2011})}\BibitemShut {NoStop}%
\bibitem [{\citenamefont {Fan}\ \emph {et~al.}(2014)\citenamefont {Fan},
  \citenamefont {Iwashita},\ and\ \citenamefont {Egami}}]{Fan2014}%
  \BibitemOpen
  \bibfield  {author} {\bibinfo {author} {\bibfnamefont {Y.}~\bibnamefont
  {Fan}}, \bibinfo {author} {\bibfnamefont {T.}~\bibnamefont {Iwashita}},\ and\
  \bibinfo {author} {\bibfnamefont {T.}~\bibnamefont {Egami}},\ }\bibfield
  {title} {\bibinfo {title} {{How thermally activated deformation starts in
  metallic glass}},\ }\href {https://doi.org/10.1038/ncomms6083} {\bibfield
  {journal} {\bibinfo  {journal} {Nat. Commun.}\ }\textbf {\bibinfo {volume}
  {5}},\ \bibinfo {pages} {5083} (\bibinfo {year} {2014})}\BibitemShut
  {NoStop}%
\bibitem [{\citenamefont {Ye}\ \emph {et~al.}(2010)\citenamefont {Ye},
  \citenamefont {Lu}, \citenamefont {Liu}, \citenamefont {Wang},\ and\
  \citenamefont {Yang}}]{Ye2010}%
  \BibitemOpen
  \bibfield  {author} {\bibinfo {author} {\bibfnamefont {J.~C.}\ \bibnamefont
  {Ye}}, \bibinfo {author} {\bibfnamefont {J.}~\bibnamefont {Lu}}, \bibinfo
  {author} {\bibfnamefont {C.~T.}\ \bibnamefont {Liu}}, \bibinfo {author}
  {\bibfnamefont {Q.}~\bibnamefont {Wang}},\ and\ \bibinfo {author}
  {\bibfnamefont {Y.}~\bibnamefont {Yang}},\ }\bibfield  {title} {\bibinfo
  {title} {{Atomistic free-volume zones and inelastic deformation of metallic
  glasses}},\ }\href {https://doi.org/10.1038/nmat2802} {\bibfield  {journal}
  {\bibinfo  {journal} {Nat. Mater.}\ }\textbf {\bibinfo {volume} {9}},\
  \bibinfo {pages} {619} (\bibinfo {year} {2010})}\BibitemShut {NoStop}%
\bibitem [{\citenamefont {Tong}\ and\ \citenamefont {Tanaka}(2018)}]{Tong2018}%
  \BibitemOpen
  \bibfield  {author} {\bibinfo {author} {\bibfnamefont {H.}~\bibnamefont
  {Tong}}\ and\ \bibinfo {author} {\bibfnamefont {H.}~\bibnamefont {Tanaka}},\
  }\bibfield  {title} {\bibinfo {title} {{Revealing Hidden Structural Order
  Controlling Both Fast and Slow Glassy Dynamics in Supercooled Liquids}},\
  }\href {https://doi.org/10.1103/PhysRevX.8.011041} {\bibfield  {journal}
  {\bibinfo  {journal} {Phys. Rev. X}\ }\textbf {\bibinfo {volume} {8}},\
  \bibinfo {pages} {011041} (\bibinfo {year} {2018})}\BibitemShut {NoStop}%
\bibitem [{\citenamefont {Spaepen}(2006)}]{Spaepen2006}%
  \BibitemOpen
  \bibfield  {author} {\bibinfo {author} {\bibfnamefont {F.}~\bibnamefont
  {Spaepen}},\ }\bibfield  {title} {\bibinfo {title} {{Homogeneous flow of
  metallic glasses: A free volume perspective}},\ }\href
  {https://doi.org/10.1016/j.scriptamat.2005.09.046} {\bibfield  {journal}
  {\bibinfo  {journal} {Scr. Mater.}\ }\textbf {\bibinfo {volume} {54}},\
  \bibinfo {pages} {363} (\bibinfo {year} {2006})}\BibitemShut {NoStop}%
\bibitem [{\citenamefont {Wang}\ \emph {et~al.}(2017)\citenamefont {Wang},
  \citenamefont {Li},\ and\ \citenamefont {Xu}}]{Wang2017}%
  \BibitemOpen
  \bibfield  {author} {\bibinfo {author} {\bibfnamefont {Y.}~\bibnamefont
  {Wang}}, \bibinfo {author} {\bibfnamefont {M.}~\bibnamefont {Li}},\ and\
  \bibinfo {author} {\bibfnamefont {J.}~\bibnamefont {Xu}},\ }\bibfield
  {title} {\bibinfo {title} {{Free volume gradient effect on mechanical
  properties of metallic glasses}},\ }\href
  {https://doi.org/10.1016/j.scriptamat.2016.11.003} {\bibfield  {journal}
  {\bibinfo  {journal} {Scr. Mater.}\ }\textbf {\bibinfo {volume} {130}},\
  \bibinfo {pages} {12} (\bibinfo {year} {2017})}\BibitemShut {NoStop}%
\bibitem [{\citenamefont {Argon}(1979)}]{Argon1979}%
  \BibitemOpen
  \bibfield  {author} {\bibinfo {author} {\bibfnamefont {A.~S.}\ \bibnamefont
  {Argon}},\ }\bibfield  {title} {\bibinfo {title} {{Plastic deformation in
  metallic glasses}},\ }\href {https://doi.org/10.1016/0001-6160(79)90055-5}
  {\bibfield  {journal} {\bibinfo  {journal} {Acta Metall.}\ }\textbf {\bibinfo
  {volume} {27}},\ \bibinfo {pages} {47} (\bibinfo {year} {1979})}\BibitemShut
  {NoStop}%
\bibitem [{\citenamefont {Falk}\ and\ \citenamefont {Langer}(1998)}]{Falk1998}%
  \BibitemOpen
  \bibfield  {author} {\bibinfo {author} {\bibfnamefont {M.~L.}\ \bibnamefont
  {Falk}}\ and\ \bibinfo {author} {\bibfnamefont {J.~S.}\ \bibnamefont
  {Langer}},\ }\bibfield  {title} {\bibinfo {title} {{Dynamics of viscoplastic
  deformation in amorphous solids}},\ }\href
  {https://doi.org/10.1103/PhysRevE.57.7192} {\bibfield  {journal} {\bibinfo
  {journal} {Phys. Rev. E}\ }\textbf {\bibinfo {volume} {57}},\ \bibinfo
  {pages} {7192} (\bibinfo {year} {1998})}\BibitemShut {NoStop}%
\bibitem [{\citenamefont {Lema{\^{i}}tre}(2002)}]{Lemaitre2002}%
  \BibitemOpen
  \bibfield  {author} {\bibinfo {author} {\bibfnamefont {A.}~\bibnamefont
  {Lema{\^{i}}tre}},\ }\bibfield  {title} {\bibinfo {title} {{Rearrangements
  and Dilatancy for Sheared Dense Materials}},\ }\href
  {https://doi.org/10.1103/PhysRevLett.89.195503} {\bibfield  {journal}
  {\bibinfo  {journal} {Phys. Rev. Lett.}\ }\textbf {\bibinfo {volume} {89}},\
  \bibinfo {pages} {195503} (\bibinfo {year} {2002})}\BibitemShut {NoStop}%
\bibitem [{\citenamefont {Johnson}\ and\ \citenamefont
  {Samwer}(2005)}]{Johnson2005}%
  \BibitemOpen
  \bibfield  {author} {\bibinfo {author} {\bibfnamefont {W.~L.}\ \bibnamefont
  {Johnson}}\ and\ \bibinfo {author} {\bibfnamefont {K.}~\bibnamefont
  {Samwer}},\ }\bibfield  {title} {\bibinfo {title} {{A universal criterion for
  plastic yielding of metallic glasses with a $(T/T_{\rm{g}})^{2/3}$
  temperature dependence}},\ }\href
  {https://doi.org/10.1103/PhysRevLett.95.195501} {\bibfield  {journal}
  {\bibinfo  {journal} {Phys. Rev. Lett.}\ }\textbf {\bibinfo {volume} {95}},\
  \bibinfo {pages} {195501} (\bibinfo {year} {2005})}\BibitemShut {NoStop}%
\bibitem [{\citenamefont {Lu}\ \emph {et~al.}(2014)\citenamefont {Lu},
  \citenamefont {Jiao}, \citenamefont {Wang},\ and\ \citenamefont
  {Bai}}]{Lu2014}%
  \BibitemOpen
  \bibfield  {author} {\bibinfo {author} {\bibfnamefont {Z.}~\bibnamefont
  {Lu}}, \bibinfo {author} {\bibfnamefont {W.}~\bibnamefont {Jiao}}, \bibinfo
  {author} {\bibfnamefont {W.~H.}\ \bibnamefont {Wang}},\ and\ \bibinfo
  {author} {\bibfnamefont {H.~Y.}\ \bibnamefont {Bai}},\ }\bibfield  {title}
  {\bibinfo {title} {{Flow unit perspective on room temperature homogeneous
  plastic deformation in metallic glasses}},\ }\href
  {https://doi.org/10.1103/PhysRevLett.113.045501} {\bibfield  {journal}
  {\bibinfo  {journal} {Phys. Rev. Lett.}\ }\textbf {\bibinfo {volume} {113}},\
  \bibinfo {pages} {045501} (\bibinfo {year} {2014})}\BibitemShut {NoStop}%
\bibitem [{\citenamefont {Wang}\ and\ \citenamefont {Wang}(2019)}]{WangZ2019}%
  \BibitemOpen
  \bibfield  {author} {\bibinfo {author} {\bibfnamefont {Z.}~\bibnamefont
  {Wang}}\ and\ \bibinfo {author} {\bibfnamefont {W.~H.}\ \bibnamefont
  {Wang}},\ }\bibfield  {title} {\bibinfo {title} {{Flow units as dynamic
  defects in metallic glassy materials}},\ }\href
  {https://doi.org/10.1093/nsr/nwy084} {\bibfield  {journal} {\bibinfo
  {journal} {Natl. Sci. Rev.}\ }\textbf {\bibinfo {volume} {6}},\ \bibinfo
  {pages} {304} (\bibinfo {year} {2019})}\BibitemShut {NoStop}%
\bibitem [{\citenamefont {Ding}\ \emph {et~al.}(2014)\citenamefont {Ding},
  \citenamefont {Patinet}, \citenamefont {Falk}, \citenamefont {Cheng},\ and\
  \citenamefont {Ma}}]{Ding2014}%
  \BibitemOpen
  \bibfield  {author} {\bibinfo {author} {\bibfnamefont {J.}~\bibnamefont
  {Ding}}, \bibinfo {author} {\bibfnamefont {S.}~\bibnamefont {Patinet}},
  \bibinfo {author} {\bibfnamefont {M.~L.}\ \bibnamefont {Falk}}, \bibinfo
  {author} {\bibfnamefont {Y.}~\bibnamefont {Cheng}},\ and\ \bibinfo {author}
  {\bibfnamefont {E.}~\bibnamefont {Ma}},\ }\bibfield  {title} {\bibinfo
  {title} {{Soft spots and their structural signature in a metallic glass}},\
  }\href {https://doi.org/10.1073/pnas.1412095111} {\bibfield  {journal}
  {\bibinfo  {journal} {Proc. Natl. Acad. Sci.}\ }\textbf {\bibinfo {volume}
  {111}},\ \bibinfo {pages} {14052} (\bibinfo {year} {2014})}\BibitemShut
  {NoStop}%
\bibitem [{\citenamefont {Jiang}\ \emph {et~al.}(2008)\citenamefont {Jiang},
  \citenamefont {Ling}, \citenamefont {Meng},\ and\ \citenamefont
  {Dai}}]{Jiang2008}%
  \BibitemOpen
  \bibfield  {author} {\bibinfo {author} {\bibfnamefont {M.~Q.}\ \bibnamefont
  {Jiang}}, \bibinfo {author} {\bibfnamefont {Z.}~\bibnamefont {Ling}},
  \bibinfo {author} {\bibfnamefont {J.~X.}\ \bibnamefont {Meng}},\ and\
  \bibinfo {author} {\bibfnamefont {L.~H.}\ \bibnamefont {Dai}},\ }\bibfield
  {title} {\bibinfo {title} {{Energy dissipation in fracture of bulk metallic
  glasses via inherent competition between local softening and
  quasi-cleavage}},\ }\href {https://doi.org/10.1080/14786430701864753}
  {\bibfield  {journal} {\bibinfo  {journal} {Philos. Mag.}\ }\textbf {\bibinfo
  {volume} {88}},\ \bibinfo {pages} {407} (\bibinfo {year} {2008})}\BibitemShut
  {NoStop}%
\bibitem [{\citenamefont {Huang}\ \emph {et~al.}(2014)\citenamefont {Huang},
  \citenamefont {Ling},\ and\ \citenamefont {Dai}}]{Huang2014}%
  \BibitemOpen
  \bibfield  {author} {\bibinfo {author} {\bibfnamefont {X.}~\bibnamefont
  {Huang}}, \bibinfo {author} {\bibfnamefont {Z.}~\bibnamefont {Ling}},\ and\
  \bibinfo {author} {\bibfnamefont {L.~H.}\ \bibnamefont {Dai}},\ }\bibfield
  {title} {\bibinfo {title} {{Ductile-to-brittle transition in spallation of
  metallic glasses}},\ }\href {https://doi.org/10.1063/1.4897552} {\bibfield
  {journal} {\bibinfo  {journal} {J. Appl. Phys.}\ }\textbf {\bibinfo {volume}
  {116}},\ \bibinfo {pages} {143503} (\bibinfo {year} {2014})}\BibitemShut
  {NoStop}%
\bibitem [{\citenamefont {Pan}\ \emph {et~al.}(2008)\citenamefont {Pan},
  \citenamefont {Inoue}, \citenamefont {Sakurai},\ and\ \citenamefont
  {Chen}}]{Pan2008}%
  \BibitemOpen
  \bibfield  {author} {\bibinfo {author} {\bibfnamefont {D.}~\bibnamefont
  {Pan}}, \bibinfo {author} {\bibfnamefont {A.}~\bibnamefont {Inoue}}, \bibinfo
  {author} {\bibfnamefont {T.}~\bibnamefont {Sakurai}},\ and\ \bibinfo {author}
  {\bibfnamefont {M.~W.}\ \bibnamefont {Chen}},\ }\bibfield  {title} {\bibinfo
  {title} {{Experimental characterization of shear transformation zones for
  plastic flow of bulk metallic glasses}},\ }\href
  {https://doi.org/10.1073/pnas.0806051105} {\bibfield  {journal} {\bibinfo
  {journal} {Proc. Natl. Acad. Sci.}\ }\textbf {\bibinfo {volume} {105}},\
  \bibinfo {pages} {14769} (\bibinfo {year} {2008})}\BibitemShut {NoStop}%
\bibitem [{\citenamefont {Manning}\ \emph {et~al.}(2007)\citenamefont
  {Manning}, \citenamefont {Langer},\ and\ \citenamefont
  {Carlson}}]{Manning2007}%
  \BibitemOpen
  \bibfield  {author} {\bibinfo {author} {\bibfnamefont {M.~L.}\ \bibnamefont
  {Manning}}, \bibinfo {author} {\bibfnamefont {J.~S.}\ \bibnamefont
  {Langer}},\ and\ \bibinfo {author} {\bibfnamefont {J.~M.}\ \bibnamefont
  {Carlson}},\ }\bibfield  {title} {\bibinfo {title} {{Strain localization in a
  shear transformation zone model for amorphous solids}},\ }\href
  {https://doi.org/10.1103/PhysRevE.76.056106} {\bibfield  {journal} {\bibinfo
  {journal} {Phys. Rev. E}\ }\textbf {\bibinfo {volume} {76}},\ \bibinfo
  {pages} {056106} (\bibinfo {year} {2007})}\BibitemShut {NoStop}%
\bibitem [{\citenamefont {Lewandowski}\ and\ \citenamefont
  {Greer}(2006)}]{Lewandowski2006}%
  \BibitemOpen
  \bibfield  {author} {\bibinfo {author} {\bibfnamefont {J.~J.}\ \bibnamefont
  {Lewandowski}}\ and\ \bibinfo {author} {\bibfnamefont {A.~L.}\ \bibnamefont
  {Greer}},\ }\bibfield  {title} {\bibinfo {title} {{Temperature rise at shear
  bands in metallic glasses}},\ }\href {https://doi.org/10.1038/nmat1536}
  {\bibfield  {journal} {\bibinfo  {journal} {Nat. Mater.}\ }\textbf {\bibinfo
  {volume} {5}},\ \bibinfo {pages} {15} (\bibinfo {year} {2006})}\BibitemShut
  {NoStop}%
\bibitem [{\citenamefont {Schmidt}\ \emph {et~al.}(2015)\citenamefont
  {Schmidt}, \citenamefont {R{\"{o}}sner}, \citenamefont {Peterlechner},
  \citenamefont {Wilde},\ and\ \citenamefont {Voyles}}]{Schmidt2015}%
  \BibitemOpen
  \bibfield  {author} {\bibinfo {author} {\bibfnamefont {V.}~\bibnamefont
  {Schmidt}}, \bibinfo {author} {\bibfnamefont {H.}~\bibnamefont
  {R{\"{o}}sner}}, \bibinfo {author} {\bibfnamefont {M.}~\bibnamefont
  {Peterlechner}}, \bibinfo {author} {\bibfnamefont {G.}~\bibnamefont
  {Wilde}},\ and\ \bibinfo {author} {\bibfnamefont {P.~M.}\ \bibnamefont
  {Voyles}},\ }\bibfield  {title} {\bibinfo {title} {{Quantitative Measurement
  of Density in a Shear Band of Metallic Glass Monitored Along its Propagation
  Direction}},\ }\href {https://doi.org/10.1103/PhysRevLett.115.035501}
  {\bibfield  {journal} {\bibinfo  {journal} {Phys. Rev. Lett.}\ }\textbf
  {\bibinfo {volume} {115}},\ \bibinfo {pages} {035501} (\bibinfo {year}
  {2015})}\BibitemShut {NoStop}%
\bibitem [{\citenamefont {Shen}\ \emph {et~al.}(2018)\citenamefont {Shen},
  \citenamefont {Luo}, \citenamefont {Hu}, \citenamefont {Bai}, \citenamefont
  {Sun}, \citenamefont {Sun}, \citenamefont {Liu},\ and\ \citenamefont
  {Wang}}]{Shen2018}%
  \BibitemOpen
  \bibfield  {author} {\bibinfo {author} {\bibfnamefont {L.~Q.}\ \bibnamefont
  {Shen}}, \bibinfo {author} {\bibfnamefont {P.}~\bibnamefont {Luo}}, \bibinfo
  {author} {\bibfnamefont {Y.~C.}\ \bibnamefont {Hu}}, \bibinfo {author}
  {\bibfnamefont {H.~Y.}\ \bibnamefont {Bai}}, \bibinfo {author} {\bibfnamefont
  {Y.~H.}\ \bibnamefont {Sun}}, \bibinfo {author} {\bibfnamefont {B.~A.}\
  \bibnamefont {Sun}}, \bibinfo {author} {\bibfnamefont {Y.~H.}\ \bibnamefont
  {Liu}},\ and\ \bibinfo {author} {\bibfnamefont {W.~H.}\ \bibnamefont
  {Wang}},\ }\bibfield  {title} {\bibinfo {title} {{Shear-band affected zone
  revealed by magnetic domains in a ferromagnetic metallic glass}},\ }\href
  {https://doi.org/10.1038/s41467-018-06919-2} {\bibfield  {journal} {\bibinfo
  {journal} {Nat. Commun.}\ }\textbf {\bibinfo {volume} {9}},\ \bibinfo {pages}
  {4414} (\bibinfo {year} {2018})}\BibitemShut {NoStop}%
\bibitem [{\citenamefont {Huang}\ \emph {et~al.}(2002)\citenamefont {Huang},
  \citenamefont {Suo}, \citenamefont {Prevost},\ and\ \citenamefont
  {Nix}}]{Huang2002}%
  \BibitemOpen
  \bibfield  {author} {\bibinfo {author} {\bibfnamefont {R.}~\bibnamefont
  {Huang}}, \bibinfo {author} {\bibfnamefont {Z.}~\bibnamefont {Suo}}, \bibinfo
  {author} {\bibfnamefont {J.~H.}\ \bibnamefont {Prevost}},\ and\ \bibinfo
  {author} {\bibfnamefont {W.~D.}\ \bibnamefont {Nix}},\ }\bibfield  {title}
  {\bibinfo {title} {{Inhomogeneous deformation in metallic glasses}},\ }\href
  {https://doi.org/10.1016/S0022-5096(01)00115-6} {\bibfield  {journal}
  {\bibinfo  {journal} {J. Mech. Phys. Solids}\ }\textbf {\bibinfo {volume}
  {50}},\ \bibinfo {pages} {1011} (\bibinfo {year} {2002})}\BibitemShut
  {NoStop}%
\bibitem [{\citenamefont {Dai}\ \emph {et~al.}(2005)\citenamefont {Dai},
  \citenamefont {Yan}, \citenamefont {Liu},\ and\ \citenamefont
  {Bai}}]{Dai2005}%
  \BibitemOpen
  \bibfield  {author} {\bibinfo {author} {\bibfnamefont {L.~H.}\ \bibnamefont
  {Dai}}, \bibinfo {author} {\bibfnamefont {M.}~\bibnamefont {Yan}}, \bibinfo
  {author} {\bibfnamefont {L.~F.}\ \bibnamefont {Liu}},\ and\ \bibinfo {author}
  {\bibfnamefont {Y.~L.}\ \bibnamefont {Bai}},\ }\bibfield  {title} {\bibinfo
  {title} {{Adiabatic shear banding instability in bulk metallic glasses}},\
  }\href {https://doi.org/10.1063/1.2067691} {\bibfield  {journal} {\bibinfo
  {journal} {Appl. Phys. Lett.}\ }\textbf {\bibinfo {volume} {87}},\ \bibinfo
  {pages} {141916} (\bibinfo {year} {2005})}\BibitemShut {NoStop}%
\bibitem [{\citenamefont {Jiang}\ and\ \citenamefont {Dai}(2009)}]{Jiang2009}%
  \BibitemOpen
  \bibfield  {author} {\bibinfo {author} {\bibfnamefont {M.~Q.}\ \bibnamefont
  {Jiang}}\ and\ \bibinfo {author} {\bibfnamefont {L.~H.}\ \bibnamefont
  {Dai}},\ }\bibfield  {title} {\bibinfo {title} {{On the origin of shear
  banding instability in metallic glasses}},\ }\href
  {https://doi.org/10.1016/j.jmps.2009.04.008} {\bibfield  {journal} {\bibinfo
  {journal} {J. Mech. Phys. Solids}\ }\textbf {\bibinfo {volume} {57}},\
  \bibinfo {pages} {1267} (\bibinfo {year} {2009})}\BibitemShut {NoStop}%
\bibitem [{\citenamefont {Han}\ \emph {et~al.}(2009)\citenamefont {Han},
  \citenamefont {Wu}, \citenamefont {Li}, \citenamefont {Wei},\ and\
  \citenamefont {Gao}}]{Han2009}%
  \BibitemOpen
  \bibfield  {author} {\bibinfo {author} {\bibfnamefont {Z.}~\bibnamefont
  {Han}}, \bibinfo {author} {\bibfnamefont {W.}~\bibnamefont {Wu}}, \bibinfo
  {author} {\bibfnamefont {Y.}~\bibnamefont {Li}}, \bibinfo {author}
  {\bibfnamefont {Y.}~\bibnamefont {Wei}},\ and\ \bibinfo {author}
  {\bibfnamefont {H.}~\bibnamefont {Gao}},\ }\bibfield  {title} {\bibinfo
  {title} {{An instability index of shear band for plasticity in metallic
  glasses}},\ }\href {https://doi.org/10.1016/j.actamat.2008.11.018} {\bibfield
   {journal} {\bibinfo  {journal} {Acta Mater.}\ }\textbf {\bibinfo {volume}
  {57}},\ \bibinfo {pages} {1367} (\bibinfo {year} {2009})}\BibitemShut
  {NoStop}%
\bibitem [{\citenamefont {Shimizu}\ \emph {et~al.}(2006)\citenamefont
  {Shimizu}, \citenamefont {Ogata},\ and\ \citenamefont {Li}}]{Shimizu2006}%
  \BibitemOpen
  \bibfield  {author} {\bibinfo {author} {\bibfnamefont {F.}~\bibnamefont
  {Shimizu}}, \bibinfo {author} {\bibfnamefont {S.}~\bibnamefont {Ogata}},\
  and\ \bibinfo {author} {\bibfnamefont {J.}~\bibnamefont {Li}},\ }\bibfield
  {title} {\bibinfo {title} {{Yield point of metallic glass}},\ }\href
  {https://doi.org/10.1016/j.actamat.2006.05.024} {\bibfield  {journal}
  {\bibinfo  {journal} {Acta Mater.}\ }\textbf {\bibinfo {volume} {54}},\
  \bibinfo {pages} {4293} (\bibinfo {year} {2006})}\BibitemShut {NoStop}%
\bibitem [{\citenamefont {Cao}\ \emph {et~al.}(2009)\citenamefont {Cao},
  \citenamefont {Cheng},\ and\ \citenamefont {Ma}}]{Cao2009}%
  \BibitemOpen
  \bibfield  {author} {\bibinfo {author} {\bibfnamefont {A.~J.}\ \bibnamefont
  {Cao}}, \bibinfo {author} {\bibfnamefont {Y.~Q.}\ \bibnamefont {Cheng}},\
  and\ \bibinfo {author} {\bibfnamefont {E.}~\bibnamefont {Ma}},\ }\bibfield
  {title} {\bibinfo {title} {{Structural processes that initiate shear
  localization in metallic glass}},\ }\href
  {https://doi.org/10.1016/j.actamat.2009.07.016} {\bibfield  {journal}
  {\bibinfo  {journal} {Acta Mater.}\ }\textbf {\bibinfo {volume} {57}},\
  \bibinfo {pages} {5146} (\bibinfo {year} {2009})}\BibitemShut {NoStop}%
\bibitem [{\citenamefont {Li}\ and\ \citenamefont {Li}(2007)}]{Li2007}%
  \BibitemOpen
  \bibfield  {author} {\bibinfo {author} {\bibfnamefont {Q.~K.}\ \bibnamefont
  {Li}}\ and\ \bibinfo {author} {\bibfnamefont {M.}~\bibnamefont {Li}},\
  }\bibfield  {title} {\bibinfo {title} {{Assessing the critical sizes for
  shear band formation in metallic glasses from molecular dynamics
  simulation}},\ }\href {https://doi.org/10.1063/1.2821832} {\bibfield
  {journal} {\bibinfo  {journal} {Appl. Phys. Lett.}\ }\textbf {\bibinfo
  {volume} {91}},\ \bibinfo {pages} {231905} (\bibinfo {year}
  {2007})}\BibitemShut {NoStop}%
\bibitem [{\citenamefont {Klaum{\"{u}}nzer}\ \emph {et~al.}(2011)\citenamefont
  {Klaum{\"{u}}nzer}, \citenamefont {Lazarev}, \citenamefont {Maa{\ss}},
  \citenamefont {{Dalla Torre}}, \citenamefont {Vinogradov},\ and\
  \citenamefont {L{\"{o}}ffler}}]{Klaumunzer2011}%
  \BibitemOpen
  \bibfield  {author} {\bibinfo {author} {\bibfnamefont {D.}~\bibnamefont
  {Klaum{\"{u}}nzer}}, \bibinfo {author} {\bibfnamefont {A.}~\bibnamefont
  {Lazarev}}, \bibinfo {author} {\bibfnamefont {R.}~\bibnamefont {Maa{\ss}}},
  \bibinfo {author} {\bibfnamefont {F.~H.}\ \bibnamefont {{Dalla Torre}}},
  \bibinfo {author} {\bibfnamefont {A.}~\bibnamefont {Vinogradov}},\ and\
  \bibinfo {author} {\bibfnamefont {J.~F.}\ \bibnamefont {L{\"{o}}ffler}},\
  }\bibfield  {title} {\bibinfo {title} {{Probing Shear-Band Initiation in
  Metallic Glasses}},\ }\href {https://doi.org/10.1103/PhysRevLett.107.185502}
  {\bibfield  {journal} {\bibinfo  {journal} {Phys. Rev. Lett.}\ }\textbf
  {\bibinfo {volume} {107}},\ \bibinfo {pages} {185502} (\bibinfo {year}
  {2011})}\BibitemShut {NoStop}%
\bibitem [{\citenamefont {Zeng}\ \emph {et~al.}(2018)\citenamefont {Zeng},
  \citenamefont {Jiang},\ and\ \citenamefont {Dai}}]{Zeng2018}%
  \BibitemOpen
  \bibfield  {author} {\bibinfo {author} {\bibfnamefont {F.}~\bibnamefont
  {Zeng}}, \bibinfo {author} {\bibfnamefont {M.~Q.}\ \bibnamefont {Jiang}},\
  and\ \bibinfo {author} {\bibfnamefont {L.~H.}\ \bibnamefont {Dai}},\
  }\bibfield  {title} {\bibinfo {title} {{Dilatancy induced ductile–brittle
  transition of shear band in metallic glasses}},\ }\href
  {https://doi.org/10.1098/rspa.2017.0836} {\bibfield  {journal} {\bibinfo
  {journal} {Proc. R. Soc. A Math. Phys. Eng. Sci.}\ }\textbf {\bibinfo
  {volume} {474}},\ \bibinfo {pages} {20170836} (\bibinfo {year}
  {2018})}\BibitemShut {NoStop}%
\bibitem [{\citenamefont {Maloney}\ and\ \citenamefont
  {Lema{\^{i}}tre}(2006)}]{Maloney2006}%
  \BibitemOpen
  \bibfield  {author} {\bibinfo {author} {\bibfnamefont {C.~E.}\ \bibnamefont
  {Maloney}}\ and\ \bibinfo {author} {\bibfnamefont {A.}~\bibnamefont
  {Lema{\^{i}}tre}},\ }\bibfield  {title} {\bibinfo {title} {{Amorphous systems
  in athermal, quasistatic shear}},\ }\href
  {https://doi.org/10.1103/PhysRevE.74.016118} {\bibfield  {journal} {\bibinfo
  {journal} {Phys. Rev. E}\ }\textbf {\bibinfo {volume} {74}},\ \bibinfo
  {pages} {016118} (\bibinfo {year} {2006})}\BibitemShut {NoStop}%
\bibitem [{\citenamefont {Hassani}\ \emph {et~al.}(2019)\citenamefont
  {Hassani}, \citenamefont {Lagogianni},\ and\ \citenamefont
  {Varnik}}]{Hassani2019}%
  \BibitemOpen
  \bibfield  {author} {\bibinfo {author} {\bibfnamefont {M.}~\bibnamefont
  {Hassani}}, \bibinfo {author} {\bibfnamefont {A.~E.}\ \bibnamefont
  {Lagogianni}},\ and\ \bibinfo {author} {\bibfnamefont {F.}~\bibnamefont
  {Varnik}},\ }\bibfield  {title} {\bibinfo {title} {{Probing the degree of
  heterogeneity within a shear band of a model glass}},\ }\href
  {https://doi.org/10.1103/PhysRevLett.123.195502} {\bibfield  {journal}
  {\bibinfo  {journal} {Phys. Rev. Lett.}\ }\textbf {\bibinfo {volume} {123}},\
  \bibinfo {pages} {195502} (\bibinfo {year} {2019})}\BibitemShut {NoStop}%
\bibitem [{\citenamefont {Sopu}\ \emph {et~al.}(2017)\citenamefont {Sopu},
  \citenamefont {Stukowski}, \citenamefont {Stoica},\ and\ \citenamefont
  {Scudino}}]{Sopu2017}%
  \BibitemOpen
  \bibfield  {author} {\bibinfo {author} {\bibfnamefont {D.}~\bibnamefont
  {Sopu}}, \bibinfo {author} {\bibfnamefont {A.}~\bibnamefont {Stukowski}},
  \bibinfo {author} {\bibfnamefont {M.}~\bibnamefont {Stoica}},\ and\ \bibinfo
  {author} {\bibfnamefont {S.}~\bibnamefont {Scudino}},\ }\bibfield  {title}
  {\bibinfo {title} {{Atomic-Level Processes of Shear Band Nucleation in
  Metallic Glasses}},\ }\href {https://doi.org/10.1103/PhysRevLett.119.195503}
  {\bibfield  {journal} {\bibinfo  {journal} {Phys. Rev. Lett.}\ }\textbf
  {\bibinfo {volume} {119}},\ \bibinfo {pages} {195503} (\bibinfo {year}
  {2017})}\BibitemShut {NoStop}%
\bibitem [{\citenamefont {Shimizu}\ \emph {et~al.}(2007)\citenamefont
  {Shimizu}, \citenamefont {Ogata},\ and\ \citenamefont {Li}}]{Shimizu2007}%
  \BibitemOpen
  \bibfield  {author} {\bibinfo {author} {\bibfnamefont {F.}~\bibnamefont
  {Shimizu}}, \bibinfo {author} {\bibfnamefont {S.}~\bibnamefont {Ogata}},\
  and\ \bibinfo {author} {\bibfnamefont {J.}~\bibnamefont {Li}},\ }\bibfield
  {title} {\bibinfo {title} {{Theory of shear banding in metallic glasses and
  molecular dynamics calculations}},\ }\href
  {https://doi.org/10.2320/matertrans.MJ200769} {\bibfield  {journal} {\bibinfo
   {journal} {Mater. Trans.}\ }\textbf {\bibinfo {volume} {48}},\ \bibinfo
  {pages} {2923} (\bibinfo {year} {2007})}\BibitemShut {NoStop}%
\bibitem [{\citenamefont {Richard}\ \emph {et~al.}(2020)\citenamefont
  {Richard}, \citenamefont {Ozawa}, \citenamefont {Patinet}, \citenamefont
  {Stanifer}, \citenamefont {Shang}, \citenamefont {Ridout}, \citenamefont
  {Xu}, \citenamefont {Zhang}, \citenamefont {Morse}, \citenamefont {Barrat},
  \citenamefont {Berthier}, \citenamefont {Falk}, \citenamefont {Guan},
  \citenamefont {Liu}, \citenamefont {Martens}, \citenamefont {Sastry},
  \citenamefont {Vandembroucq}, \citenamefont {Lerner},\ and\ \citenamefont
  {Manning}}]{Richard2020}%
  \BibitemOpen
  \bibfield  {author} {\bibinfo {author} {\bibfnamefont {D.}~\bibnamefont
  {Richard}}, \bibinfo {author} {\bibfnamefont {M.}~\bibnamefont {Ozawa}},
  \bibinfo {author} {\bibfnamefont {S.}~\bibnamefont {Patinet}}, \bibinfo
  {author} {\bibfnamefont {E.}~\bibnamefont {Stanifer}}, \bibinfo {author}
  {\bibfnamefont {B.}~\bibnamefont {Shang}}, \bibinfo {author} {\bibfnamefont
  {S.~A.}\ \bibnamefont {Ridout}}, \bibinfo {author} {\bibfnamefont
  {B.}~\bibnamefont {Xu}}, \bibinfo {author} {\bibfnamefont {G.}~\bibnamefont
  {Zhang}}, \bibinfo {author} {\bibfnamefont {P.~K.}\ \bibnamefont {Morse}},
  \bibinfo {author} {\bibfnamefont {J.-L.}\ \bibnamefont {Barrat}}, \bibinfo
  {author} {\bibfnamefont {L.}~\bibnamefont {Berthier}}, \bibinfo {author}
  {\bibfnamefont {M.~L.}\ \bibnamefont {Falk}}, \bibinfo {author}
  {\bibfnamefont {P.}~\bibnamefont {Guan}}, \bibinfo {author} {\bibfnamefont
  {A.~J.}\ \bibnamefont {Liu}}, \bibinfo {author} {\bibfnamefont
  {K.}~\bibnamefont {Martens}}, \bibinfo {author} {\bibfnamefont
  {S.}~\bibnamefont {Sastry}}, \bibinfo {author} {\bibfnamefont
  {D.}~\bibnamefont {Vandembroucq}}, \bibinfo {author} {\bibfnamefont
  {E.}~\bibnamefont {Lerner}},\ and\ \bibinfo {author} {\bibfnamefont {M.~L.}\
  \bibnamefont {Manning}},\ }\bibfield  {title} {\bibinfo {title} {{Predicting
  plasticity in disordered solids from structural indicators}},\ }\href
  {https://doi.org/10.1103/PhysRevMaterials.4.113609} {\bibfield  {journal}
  {\bibinfo  {journal} {Phys. Rev. Mater.}\ }\textbf {\bibinfo {volume} {4}},\
  \bibinfo {pages} {113609} (\bibinfo {year} {2020})}\BibitemShut {NoStop}%
\bibitem [{\citenamefont {Zhang}\ \emph {et~al.}(2021)\citenamefont {Zhang},
  \citenamefont {Ridout},\ and\ \citenamefont {Liu}}]{Zhang2021}%
  \BibitemOpen
  \bibfield  {author} {\bibinfo {author} {\bibfnamefont {G.}~\bibnamefont
  {Zhang}}, \bibinfo {author} {\bibfnamefont {S.~A.}\ \bibnamefont {Ridout}},\
  and\ \bibinfo {author} {\bibfnamefont {A.~J.}\ \bibnamefont {Liu}},\
  }\bibfield  {title} {\bibinfo {title} {{Interplay of Rearrangements, Strain,
  and Local Structure during Avalanche Propagation}},\ }\href
  {https://doi.org/10.1103/PhysRevX.11.041019} {\bibfield  {journal} {\bibinfo
  {journal} {Phys. Rev. X}\ }\textbf {\bibinfo {volume} {11}},\ \bibinfo
  {pages} {041019} (\bibinfo {year} {2021})}\BibitemShut {NoStop}%
\bibitem [{\citenamefont {Plimpton}(1995)}]{Plimpton1995}%
  \BibitemOpen
  \bibfield  {author} {\bibinfo {author} {\bibfnamefont {S.}~\bibnamefont
  {Plimpton}},\ }\bibfield  {title} {\bibinfo {title} {{Fast Parallel
  Algorithms for Short-Range Molecular Dynamics}},\ }\href
  {https://doi.org/10.1006/jcph.1995.1039} {\bibfield  {journal} {\bibinfo
  {journal} {J. Comput. Phys.}\ }\textbf {\bibinfo {volume} {117}},\ \bibinfo
  {pages} {1} (\bibinfo {year} {1995})}\BibitemShut {NoStop}%
\bibitem [{\citenamefont {Cheng}\ \emph {et~al.}(2009)\citenamefont {Cheng},
  \citenamefont {Ma},\ and\ \citenamefont {Sheng}}]{Cheng2009}%
  \BibitemOpen
  \bibfield  {author} {\bibinfo {author} {\bibfnamefont {Y.~Q.}\ \bibnamefont
  {Cheng}}, \bibinfo {author} {\bibfnamefont {E.}~\bibnamefont {Ma}},\ and\
  \bibinfo {author} {\bibfnamefont {H.~W.}\ \bibnamefont {Sheng}},\ }\bibfield
  {title} {\bibinfo {title} {{Atomic Level Structure in Multicomponent Bulk
  Metallic Glass}},\ }\href {https://doi.org/10.1103/PhysRevLett.102.245501}
  {\bibfield  {journal} {\bibinfo  {journal} {Phys. Rev. Lett.}\ }\textbf
  {\bibinfo {volume} {102}},\ \bibinfo {pages} {245501} (\bibinfo {year}
  {2009})}\BibitemShut {NoStop}%
\bibitem [{\citenamefont {Parrinello}\ and\ \citenamefont
  {Rahman}(1981)}]{Parrinello1981}%
  \BibitemOpen
  \bibfield  {author} {\bibinfo {author} {\bibfnamefont {M.}~\bibnamefont
  {Parrinello}}\ and\ \bibinfo {author} {\bibfnamefont {A.}~\bibnamefont
  {Rahman}},\ }\bibfield  {title} {\bibinfo {title} {{Polymorphic transitions
  in single crystals: A new molecular dynamics method}},\ }\href
  {https://doi.org/10.1063/1.328693} {\bibfield  {journal} {\bibinfo  {journal}
  {J. Appl. Phys.}\ }\textbf {\bibinfo {volume} {52}},\ \bibinfo {pages} {7182}
  (\bibinfo {year} {1981})}\BibitemShut {NoStop}%
\bibitem [{\citenamefont {Nos{\'{e}}}(1984)}]{Nose1984}%
  \BibitemOpen
  \bibfield  {author} {\bibinfo {author} {\bibfnamefont {S.}~\bibnamefont
  {Nos{\'{e}}}},\ }\bibfield  {title} {\bibinfo {title} {{A unified formulation
  of the constant temperature molecular dynamics methods}},\ }\href
  {https://doi.org/10.1063/1.447334} {\bibfield  {journal} {\bibinfo  {journal}
  {J. Chem. Phys.}\ }\textbf {\bibinfo {volume} {81}},\ \bibinfo {pages} {511}
  (\bibinfo {year} {1984})}\BibitemShut {NoStop}%
\bibitem [{\citenamefont {Hoover}(1985)}]{Hoover1985}%
  \BibitemOpen
  \bibfield  {author} {\bibinfo {author} {\bibfnamefont {W.~G.}\ \bibnamefont
  {Hoover}},\ }\bibfield  {title} {\bibinfo {title} {{Canonical dynamics:
  Equilibrium phase-space distributions}},\ }\href
  {https://doi.org/10.1103/PhysRevA.31.1695} {\bibfield  {journal} {\bibinfo
  {journal} {Phys. Rev. A}\ }\textbf {\bibinfo {volume} {31}},\ \bibinfo
  {pages} {1695} (\bibinfo {year} {1985})}\BibitemShut {NoStop}%
\bibitem [{\citenamefont {Peng}\ \emph {et~al.}(2011)\citenamefont {Peng},
  \citenamefont {Li},\ and\ \citenamefont {Wang}}]{Peng2011}%
  \BibitemOpen
  \bibfield  {author} {\bibinfo {author} {\bibfnamefont {H.~L.}\ \bibnamefont
  {Peng}}, \bibinfo {author} {\bibfnamefont {M.~Z.}\ \bibnamefont {Li}},\ and\
  \bibinfo {author} {\bibfnamefont {W.~H.}\ \bibnamefont {Wang}},\ }\bibfield
  {title} {\bibinfo {title} {{Structural Signature of Plastic Deformation in
  Metallic Glasses}},\ }\href {https://doi.org/10.1103/PhysRevLett.106.135503}
  {\bibfield  {journal} {\bibinfo  {journal} {Phys. Rev. Lett.}\ }\textbf
  {\bibinfo {volume} {106}},\ \bibinfo {pages} {135503} (\bibinfo {year}
  {2011})}\BibitemShut {NoStop}%
\bibitem [{\citenamefont {Cubuk}\ \emph {et~al.}(2017)\citenamefont {Cubuk},
  \citenamefont {Ivancic}, \citenamefont {Schoenholz}, \citenamefont
  {Strickland}, \citenamefont {Basu}, \citenamefont {Davidson}, \citenamefont
  {Fontaine}, \citenamefont {Hor}, \citenamefont {Huang}, \citenamefont
  {Jiang}, \citenamefont {Keim}, \citenamefont {Koshigan}, \citenamefont
  {Lefever}, \citenamefont {Liu}, \citenamefont {Ma}, \citenamefont
  {Magagnosc}, \citenamefont {Morrow}, \citenamefont {Ortiz}, \citenamefont
  {Rieser}, \citenamefont {Shavit}, \citenamefont {Still}, \citenamefont {Xu},
  \citenamefont {Zhang}, \citenamefont {Nordstrom}, \citenamefont {Arratia},
  \citenamefont {Carpick}, \citenamefont {Durian}, \citenamefont {Fakhraai},
  \citenamefont {Jerolmack}, \citenamefont {Lee}, \citenamefont {Li},
  \citenamefont {Riggleman}, \citenamefont {Turner}, \citenamefont {Yodh},
  \citenamefont {Gianola},\ and\ \citenamefont {Liu}}]{Cubuk2017}%
  \BibitemOpen
  \bibfield  {author} {\bibinfo {author} {\bibfnamefont {E.~D.}\ \bibnamefont
  {Cubuk}}, \bibinfo {author} {\bibfnamefont {R.~J.~S.}\ \bibnamefont
  {Ivancic}}, \bibinfo {author} {\bibfnamefont {S.~S.}\ \bibnamefont
  {Schoenholz}}, \bibinfo {author} {\bibfnamefont {D.~J.}\ \bibnamefont
  {Strickland}}, \bibinfo {author} {\bibfnamefont {A.}~\bibnamefont {Basu}},
  \bibinfo {author} {\bibfnamefont {Z.~S.}\ \bibnamefont {Davidson}}, \bibinfo
  {author} {\bibfnamefont {J.}~\bibnamefont {Fontaine}}, \bibinfo {author}
  {\bibfnamefont {J.~L.}\ \bibnamefont {Hor}}, \bibinfo {author} {\bibfnamefont
  {Y.-R.}\ \bibnamefont {Huang}}, \bibinfo {author} {\bibfnamefont
  {Y.}~\bibnamefont {Jiang}}, \bibinfo {author} {\bibfnamefont {N.~C.}\
  \bibnamefont {Keim}}, \bibinfo {author} {\bibfnamefont {K.~D.}\ \bibnamefont
  {Koshigan}}, \bibinfo {author} {\bibfnamefont {J.~A.}\ \bibnamefont
  {Lefever}}, \bibinfo {author} {\bibfnamefont {T.}~\bibnamefont {Liu}},
  \bibinfo {author} {\bibfnamefont {X.-G.}\ \bibnamefont {Ma}}, \bibinfo
  {author} {\bibfnamefont {D.~J.}\ \bibnamefont {Magagnosc}}, \bibinfo {author}
  {\bibfnamefont {E.}~\bibnamefont {Morrow}}, \bibinfo {author} {\bibfnamefont
  {C.~P.}\ \bibnamefont {Ortiz}}, \bibinfo {author} {\bibfnamefont {J.~M.}\
  \bibnamefont {Rieser}}, \bibinfo {author} {\bibfnamefont {A.}~\bibnamefont
  {Shavit}}, \bibinfo {author} {\bibfnamefont {T.}~\bibnamefont {Still}},
  \bibinfo {author} {\bibfnamefont {Y.}~\bibnamefont {Xu}}, \bibinfo {author}
  {\bibfnamefont {Y.}~\bibnamefont {Zhang}}, \bibinfo {author} {\bibfnamefont
  {K.~N.}\ \bibnamefont {Nordstrom}}, \bibinfo {author} {\bibfnamefont {P.~E.}\
  \bibnamefont {Arratia}}, \bibinfo {author} {\bibfnamefont {R.~W.}\
  \bibnamefont {Carpick}}, \bibinfo {author} {\bibfnamefont {D.~J.}\
  \bibnamefont {Durian}}, \bibinfo {author} {\bibfnamefont {Z.}~\bibnamefont
  {Fakhraai}}, \bibinfo {author} {\bibfnamefont {D.~J.}\ \bibnamefont
  {Jerolmack}}, \bibinfo {author} {\bibfnamefont {D.}~\bibnamefont {Lee}},
  \bibinfo {author} {\bibfnamefont {J.}~\bibnamefont {Li}}, \bibinfo {author}
  {\bibfnamefont {R.}~\bibnamefont {Riggleman}}, \bibinfo {author}
  {\bibfnamefont {K.~T.}\ \bibnamefont {Turner}}, \bibinfo {author}
  {\bibfnamefont {A.~G.}\ \bibnamefont {Yodh}}, \bibinfo {author}
  {\bibfnamefont {D.~S.}\ \bibnamefont {Gianola}},\ and\ \bibinfo {author}
  {\bibfnamefont {A.~J.}\ \bibnamefont {Liu}},\ }\bibfield  {title} {\bibinfo
  {title} {{Structure-property relationships from universal signatures of
  plasticity in disordered solids}},\ }\href
  {https://doi.org/10.1126/science.aai8830} {\bibfield  {journal} {\bibinfo
  {journal} {Science}\ }\textbf {\bibinfo {volume} {358}},\ \bibinfo {pages}
  {1033} (\bibinfo {year} {2017})}\BibitemShut {NoStop}%
\bibitem [{\citenamefont {Cao}\ \emph {et~al.}(2014)\citenamefont {Cao},
  \citenamefont {Lin},\ and\ \citenamefont {Park}}]{Cao2014}%
  \BibitemOpen
  \bibfield  {author} {\bibinfo {author} {\bibfnamefont {P.}~\bibnamefont
  {Cao}}, \bibinfo {author} {\bibfnamefont {X.}~\bibnamefont {Lin}},\ and\
  \bibinfo {author} {\bibfnamefont {H.~S.}\ \bibnamefont {Park}},\ }\bibfield
  {title} {\bibinfo {title} {{Surface shear-transformation zones in amorphous
  solids}},\ }\href {https://doi.org/10.1103/PhysRevE.90.012311} {\bibfield
  {journal} {\bibinfo  {journal} {Phys. Rev. E}\ }\textbf {\bibinfo {volume}
  {90}},\ \bibinfo {pages} {012311} (\bibinfo {year} {2014})}\BibitemShut
  {NoStop}%
\bibitem [{\citenamefont {Cao}\ \emph {et~al.}(2017)\citenamefont {Cao},
  \citenamefont {Short},\ and\ \citenamefont {Yip}}]{Cao2017}%
  \BibitemOpen
  \bibfield  {author} {\bibinfo {author} {\bibfnamefont {P.}~\bibnamefont
  {Cao}}, \bibinfo {author} {\bibfnamefont {M.~P.}\ \bibnamefont {Short}},\
  and\ \bibinfo {author} {\bibfnamefont {S.}~\bibnamefont {Yip}},\ }\bibfield
  {title} {\bibinfo {title} {{Understanding the mechanisms of amorphous creep
  through molecular simulation}},\ }\href
  {https://doi.org/10.1073/pnas.1708618114} {\bibfield  {journal} {\bibinfo
  {journal} {Proc. Natl. Acad. Sci.}\ }\textbf {\bibinfo {volume} {114}},\
  \bibinfo {pages} {13631} (\bibinfo {year} {2017})}\BibitemShut {NoStop}%
\bibitem [{\citenamefont {Wang}\ \emph {et~al.}(2018)\citenamefont {Wang},
  \citenamefont {Ding}, \citenamefont {Yan}, \citenamefont {Asta},
  \citenamefont {Ritchie},\ and\ \citenamefont {Li}}]{Wang2018}%
  \BibitemOpen
  \bibfield  {author} {\bibinfo {author} {\bibfnamefont {N.}~\bibnamefont
  {Wang}}, \bibinfo {author} {\bibfnamefont {J.}~\bibnamefont {Ding}}, \bibinfo
  {author} {\bibfnamefont {F.}~\bibnamefont {Yan}}, \bibinfo {author}
  {\bibfnamefont {M.}~\bibnamefont {Asta}}, \bibinfo {author} {\bibfnamefont
  {R.~O.}\ \bibnamefont {Ritchie}},\ and\ \bibinfo {author} {\bibfnamefont
  {L.}~\bibnamefont {Li}},\ }\bibfield  {title} {\bibinfo {title} {{Spatial
  correlation of elastic heterogeneity tunes the deformation behavior of
  metallic glasses}},\ }\href {https://doi.org/10.1038/s41524-018-0077-8}
  {\bibfield  {journal} {\bibinfo  {journal} {npj Comput. Mater.}\ }\textbf
  {\bibinfo {volume} {4}},\ \bibinfo {pages} {19} (\bibinfo {year}
  {2018})}\BibitemShut {NoStop}%
\bibitem [{\citenamefont {Hufnagel}\ \emph {et~al.}(2016)\citenamefont
  {Hufnagel}, \citenamefont {Schuh},\ and\ \citenamefont
  {Falk}}]{Hufnagel2016}%
  \BibitemOpen
  \bibfield  {author} {\bibinfo {author} {\bibfnamefont {T.~C.}\ \bibnamefont
  {Hufnagel}}, \bibinfo {author} {\bibfnamefont {C.~A.}\ \bibnamefont
  {Schuh}},\ and\ \bibinfo {author} {\bibfnamefont {M.~L.}\ \bibnamefont
  {Falk}},\ }\bibfield  {title} {\bibinfo {title} {{Deformation of metallic
  glasses: Recent developments in theory, simulations, and experiments}},\
  }\href {https://doi.org/10.1016/j.actamat.2016.01.049} {\bibfield  {journal}
  {\bibinfo  {journal} {Acta Mater.}\ }\textbf {\bibinfo {volume} {109}},\
  \bibinfo {pages} {375} (\bibinfo {year} {2016})}\BibitemShut {NoStop}%
\bibitem [{\citenamefont {Sha}\ \emph {et~al.}(2015)\citenamefont {Sha},
  \citenamefont {Qu}, \citenamefont {Liu}, \citenamefont {Wang},\ and\
  \citenamefont {Gao}}]{Sha2015}%
  \BibitemOpen
  \bibfield  {author} {\bibinfo {author} {\bibfnamefont {Z.~D.}\ \bibnamefont
  {Sha}}, \bibinfo {author} {\bibfnamefont {S.~X.}\ \bibnamefont {Qu}},
  \bibinfo {author} {\bibfnamefont {Z.~S.}\ \bibnamefont {Liu}}, \bibinfo
  {author} {\bibfnamefont {T.~J.}\ \bibnamefont {Wang}},\ and\ \bibinfo
  {author} {\bibfnamefont {H.}~\bibnamefont {Gao}},\ }\bibfield  {title}
  {\bibinfo {title} {{Cyclic Deformation in Metallic Glasses}},\ }\href
  {https://doi.org/10.1021/acs.nanolett.5b03045} {\bibfield  {journal}
  {\bibinfo  {journal} {Nano Lett.}\ }\textbf {\bibinfo {volume} {15}},\
  \bibinfo {pages} {7010} (\bibinfo {year} {2015})}\BibitemShut {NoStop}%
\bibitem [{\citenamefont {Gao}\ \emph {et~al.}(1999)\citenamefont {Gao},
  \citenamefont {Huang}, \citenamefont {Nix},\ and\ \citenamefont
  {Hutchinson}}]{Gao1999}%
  \BibitemOpen
  \bibfield  {author} {\bibinfo {author} {\bibfnamefont {H.}~\bibnamefont
  {Gao}}, \bibinfo {author} {\bibfnamefont {Y.}~\bibnamefont {Huang}}, \bibinfo
  {author} {\bibfnamefont {W.~D.}\ \bibnamefont {Nix}},\ and\ \bibinfo {author}
  {\bibfnamefont {J.~W.}\ \bibnamefont {Hutchinson}},\ }\bibfield  {title}
  {\bibinfo {title} {{Mechanism-based strain gradient plasticity - I.
  Theory}},\ }\href {https://doi.org/10.1016/S0022-5096(98)00103-3} {\bibfield
  {journal} {\bibinfo  {journal} {J. Mech. Phys. Solids}\ }\textbf {\bibinfo
  {volume} {47}},\ \bibinfo {pages} {1239} (\bibinfo {year}
  {1999})}\BibitemShut {NoStop}%
\bibitem [{sm()}]{sm}%
  \BibitemOpen
  \href@noop {} {}\bibinfo {note} {See the Supplemental Materials for further
  supporting figures.}\BibitemShut {Stop}%
\bibitem [{\citenamefont {Tian}\ \emph {et~al.}(2017)\citenamefont {Tian},
  \citenamefont {Wang}, \citenamefont {Chen},\ and\ \citenamefont
  {Dai}}]{Tian2017}%
  \BibitemOpen
  \bibfield  {author} {\bibinfo {author} {\bibfnamefont {Z.~L.}\ \bibnamefont
  {Tian}}, \bibinfo {author} {\bibfnamefont {Y.~J.}\ \bibnamefont {Wang}},
  \bibinfo {author} {\bibfnamefont {Y.}~\bibnamefont {Chen}},\ and\ \bibinfo
  {author} {\bibfnamefont {L.~H.}\ \bibnamefont {Dai}},\ }\bibfield  {title}
  {\bibinfo {title} {{Strain gradient drives shear banding in metallic
  glasses}},\ }\href {https://doi.org/10.1103/PhysRevB.96.094103} {\bibfield
  {journal} {\bibinfo  {journal} {Phys. Rev. B}\ }\textbf {\bibinfo {volume}
  {96}},\ \bibinfo {pages} {094103} (\bibinfo {year} {2017})}\BibitemShut
  {NoStop}%
\bibitem [{\citenamefont {Dai}\ \emph {et~al.}(2004)\citenamefont {Dai},
  \citenamefont {Liu},\ and\ \citenamefont {Bai}}]{Dai2004}%
  \BibitemOpen
  \bibfield  {author} {\bibinfo {author} {\bibfnamefont {L.~H.}\ \bibnamefont
  {Dai}}, \bibinfo {author} {\bibfnamefont {L.~F.}\ \bibnamefont {Liu}},\ and\
  \bibinfo {author} {\bibfnamefont {Y.~L.}\ \bibnamefont {Bai}},\ }\bibfield
  {title} {\bibinfo {title} {{Formation of adiabatic shear band in metal matrix
  composites}},\ }\href {https://doi.org/10.1016/j.ijsolstr.2004.05.023}
  {\bibfield  {journal} {\bibinfo  {journal} {Int. J. Solids Struct.}\ }\textbf
  {\bibinfo {volume} {41}},\ \bibinfo {pages} {5979} (\bibinfo {year}
  {2004})}\BibitemShut {NoStop}%
\bibitem [{\citenamefont {Liu}\ \emph {et~al.}(2020)\citenamefont {Liu},
  \citenamefont {Tian}, \citenamefont {Zhang}, \citenamefont {Chen},
  \citenamefont {Liu}, \citenamefont {Chen}, \citenamefont {Wang},\ and\
  \citenamefont {Dai}}]{Liu2020}%
  \BibitemOpen
  \bibfield  {author} {\bibinfo {author} {\bibfnamefont {X.~F.}\ \bibnamefont
  {Liu}}, \bibinfo {author} {\bibfnamefont {Z.~L.}\ \bibnamefont {Tian}},
  \bibinfo {author} {\bibfnamefont {X.~F.}\ \bibnamefont {Zhang}}, \bibinfo
  {author} {\bibfnamefont {H.~H.}\ \bibnamefont {Chen}}, \bibinfo {author}
  {\bibfnamefont {T.~W.}\ \bibnamefont {Liu}}, \bibinfo {author} {\bibfnamefont
  {Y.}~\bibnamefont {Chen}}, \bibinfo {author} {\bibfnamefont {Y.~J.}\
  \bibnamefont {Wang}},\ and\ \bibinfo {author} {\bibfnamefont {L.~H.}\
  \bibnamefont {Dai}},\ }\bibfield  {title} {\bibinfo {title}
  {{“Self-sharpening” tungsten high-entropy alloy}},\ }\href
  {https://doi.org/10.1016/j.actamat.2020.01.005} {\bibfield  {journal}
  {\bibinfo  {journal} {Acta Mater.}\ }\textbf {\bibinfo {volume} {186}},\
  \bibinfo {pages} {257} (\bibinfo {year} {2020})}\BibitemShut {NoStop}%
\bibitem [{\citenamefont {Fleck}\ and\ \citenamefont
  {Hutchinson}(1997)}]{Fleck1997}%
  \BibitemOpen
  \bibfield  {author} {\bibinfo {author} {\bibfnamefont {N.~A.}\ \bibnamefont
  {Fleck}}\ and\ \bibinfo {author} {\bibfnamefont {J.~W.}\ \bibnamefont
  {Hutchinson}},\ }\bibfield  {title} {\bibinfo {title} {{Strain Gradient
  Plasticity}},\ }\href {https://doi.org/10.1016/S0065-2156(08)70388-0}
  {\bibfield  {journal} {\bibinfo  {journal} {Adv. Appl. Mech.}\ }\textbf
  {\bibinfo {volume} {33}},\ \bibinfo {pages} {295} (\bibinfo {year}
  {1997})}\BibitemShut {NoStop}%
\bibitem [{\citenamefont {Spaepen}(2002)}]{Spaepen2002}%
  \BibitemOpen
  \bibfield  {author} {\bibinfo {author} {\bibfnamefont {F.}~\bibnamefont
  {Spaepen}},\ }\bibfield  {title} {\bibinfo {title} {{Five-fold symmetry in
  liquids}},\ }\href {https://doi.org/10.1038/35048652} {\bibfield  {journal}
  {\bibinfo  {journal} {Nature}\ }\textbf {\bibinfo {volume} {408}},\ \bibinfo
  {pages} {781} (\bibinfo {year} {2002})}\BibitemShut {NoStop}%
\bibitem [{\citenamefont {Hu}\ \emph {et~al.}(2015)\citenamefont {Hu},
  \citenamefont {Li}, \citenamefont {Li}, \citenamefont {Bai},\ and\
  \citenamefont {Wang}}]{Hu2015}%
  \BibitemOpen
  \bibfield  {author} {\bibinfo {author} {\bibfnamefont {Y.~C.}\ \bibnamefont
  {Hu}}, \bibinfo {author} {\bibfnamefont {F.~X.}\ \bibnamefont {Li}}, \bibinfo
  {author} {\bibfnamefont {M.~Z.}\ \bibnamefont {Li}}, \bibinfo {author}
  {\bibfnamefont {H.~Y.}\ \bibnamefont {Bai}},\ and\ \bibinfo {author}
  {\bibfnamefont {W.~H.}\ \bibnamefont {Wang}},\ }\bibfield  {title} {\bibinfo
  {title} {{Five-fold symmetry as indicator of dynamic arrest in metallic
  glass-forming liquids}},\ }\href {https://doi.org/10.1038/ncomms9310}
  {\bibfield  {journal} {\bibinfo  {journal} {Nat. Commun.}\ }\textbf {\bibinfo
  {volume} {6}},\ \bibinfo {pages} {9310} (\bibinfo {year} {2015})}\BibitemShut
  {NoStop}%
\bibitem [{\citenamefont {Zink}\ \emph {et~al.}(2006)\citenamefont {Zink},
  \citenamefont {Samwer}, \citenamefont {Johnson},\ and\ \citenamefont
  {Mayr}}]{Zink2006}%
  \BibitemOpen
  \bibfield  {author} {\bibinfo {author} {\bibfnamefont {M.}~\bibnamefont
  {Zink}}, \bibinfo {author} {\bibfnamefont {K.}~\bibnamefont {Samwer}},
  \bibinfo {author} {\bibfnamefont {W.~L.}\ \bibnamefont {Johnson}},\ and\
  \bibinfo {author} {\bibfnamefont {S.~G.}\ \bibnamefont {Mayr}},\ }\bibfield
  {title} {\bibinfo {title} {{Plastic deformation of metallic glasses: Size of
  shear transformation zones from molecular dynamics simulations}},\ }\href
  {https://doi.org/10.1103/PhysRevB.73.172203} {\bibfield  {journal} {\bibinfo
  {journal} {Phys. Rev. B}\ }\textbf {\bibinfo {volume} {73}},\ \bibinfo
  {pages} {172203} (\bibinfo {year} {2006})}\BibitemShut {NoStop}%
\bibitem [{\citenamefont {Ruan}\ \emph {et~al.}(2022)\citenamefont {Ruan},
  \citenamefont {Patinet},\ and\ \citenamefont {Falk}}]{Ruan2022}%
  \BibitemOpen
  \bibfield  {author} {\bibinfo {author} {\bibfnamefont {D.}~\bibnamefont
  {Ruan}}, \bibinfo {author} {\bibfnamefont {S.}~\bibnamefont {Patinet}},\ and\
  \bibinfo {author} {\bibfnamefont {M.~L.}\ \bibnamefont {Falk}},\ }\bibfield
  {title} {\bibinfo {title} {{Predicting plastic events and quantifying the
  local yield surface in 3D model glasses}},\ }\href
  {https://doi.org/10.1016/j.jmps.2021.104671} {\bibfield  {journal} {\bibinfo
  {journal} {J. Mech. Phys. Solids}\ }\textbf {\bibinfo {volume} {158}},\
  \bibinfo {pages} {104671} (\bibinfo {year} {2022})}\BibitemShut {NoStop}%
\bibitem [{\citenamefont {Swinney}\ and\ \citenamefont
  {Gollub}(1978)}]{Swinney1978}%
  \BibitemOpen
  \bibfield  {author} {\bibinfo {author} {\bibfnamefont {H.~L.}\ \bibnamefont
  {Swinney}}\ and\ \bibinfo {author} {\bibfnamefont {J.~P.}\ \bibnamefont
  {Gollub}},\ }\bibfield  {title} {\bibinfo {title} {{The transition to
  turbulence}},\ }\href {https://doi.org/10.1063/1.2995142} {\bibfield
  {journal} {\bibinfo  {journal} {Phys. Today}\ }\textbf {\bibinfo {volume}
  {31}},\ \bibinfo {pages} {41} (\bibinfo {year} {1978})}\BibitemShut {NoStop}%
\bibitem [{\citenamefont {Falkovich}\ and\ \citenamefont
  {Sreenivasan}(2006)}]{Falkovich2006}%
  \BibitemOpen
  \bibfield  {author} {\bibinfo {author} {\bibfnamefont {G.}~\bibnamefont
  {Falkovich}}\ and\ \bibinfo {author} {\bibfnamefont {K.~R.}\ \bibnamefont
  {Sreenivasan}},\ }\bibfield  {title} {\bibinfo {title} {{Lessons from
  hydrodynamic turbulence}},\ }\href {https://doi.org/10.1063/1.2207037}
  {\bibfield  {journal} {\bibinfo  {journal} {Phys. Today}\ }\textbf {\bibinfo
  {volume} {59}},\ \bibinfo {pages} {43} (\bibinfo {year} {2006})}\BibitemShut
  {NoStop}%
\bibitem [{\citenamefont {Ooi}\ \emph {et~al.}(1999)\citenamefont {Ooi},
  \citenamefont {Martin}, \citenamefont {Soria},\ and\ \citenamefont
  {Chong}}]{Ooi1999}%
  \BibitemOpen
  \bibfield  {author} {\bibinfo {author} {\bibfnamefont {A.}~\bibnamefont
  {Ooi}}, \bibinfo {author} {\bibfnamefont {J.}~\bibnamefont {Martin}},
  \bibinfo {author} {\bibfnamefont {J.}~\bibnamefont {Soria}},\ and\ \bibinfo
  {author} {\bibfnamefont {M.~S.}\ \bibnamefont {Chong}},\ }\bibfield  {title}
  {\bibinfo {title} {{A study of the evolution and characteristics of the
  invariants of the velocity-gradient tensor in isotropic turbulence}},\ }\href
  {https://doi.org/10.1017/S0022112098003681} {\bibfield  {journal} {\bibinfo
  {journal} {J. Fluid Mech.}\ }\textbf {\bibinfo {volume} {381}},\ \bibinfo
  {pages} {141} (\bibinfo {year} {1999})}\BibitemShut {NoStop}%
\bibitem [{\citenamefont {Castillo}(1988)}]{1988Extreme}%
  \BibitemOpen
  \bibfield  {author} {\bibinfo {author} {\bibfnamefont {E.}~\bibnamefont
  {Castillo}},\ }\href@noop {} {\emph {\bibinfo {title} {Extreme value theory
  in engineering}}}\ (\bibinfo  {publisher} {Academic Press, Inc.},\ \bibinfo
  {year} {1988})\BibitemShut {NoStop}%
\bibitem [{\citenamefont {Zhao}\ \emph {et~al.}(2013)\citenamefont {Zhao},
  \citenamefont {Li},\ and\ \citenamefont {Wang}}]{Zhao2013}%
  \BibitemOpen
  \bibfield  {author} {\bibinfo {author} {\bibfnamefont {P.}~\bibnamefont
  {Zhao}}, \bibinfo {author} {\bibfnamefont {J.}~\bibnamefont {Li}},\ and\
  \bibinfo {author} {\bibfnamefont {Y.}~\bibnamefont {Wang}},\ }\bibfield
  {title} {\bibinfo {title} {{Heterogeneously randomized STZ model of metallic
  glasses: Softening and extreme value statistics during deformation}},\ }\href
  {https://doi.org/10.1016/j.ijplas.2012.06.007} {\bibfield  {journal}
  {\bibinfo  {journal} {Int. J. Plast.}\ }\textbf {\bibinfo {volume} {40}},\
  \bibinfo {pages} {1} (\bibinfo {year} {2013})}\BibitemShut {NoStop}%
\bibitem [{\citenamefont {Cao}\ \emph {et~al.}(2019)\citenamefont {Cao},
  \citenamefont {Short},\ and\ \citenamefont {Yip}}]{Cao2019}%
  \BibitemOpen
  \bibfield  {author} {\bibinfo {author} {\bibfnamefont {P.}~\bibnamefont
  {Cao}}, \bibinfo {author} {\bibfnamefont {M.~P.}\ \bibnamefont {Short}},\
  and\ \bibinfo {author} {\bibfnamefont {S.}~\bibnamefont {Yip}},\ }\bibfield
  {title} {\bibinfo {title} {{Potential energy landscape activations governing
  plastic flows in glass rheology}},\ }\href
  {https://doi.org/10.1073/pnas.1907317116} {\bibfield  {journal} {\bibinfo
  {journal} {Proc. Natl. Acad. Sci.}\ }\textbf {\bibinfo {volume} {116}},\
  \bibinfo {pages} {18790} (\bibinfo {year} {2019})}\BibitemShut {NoStop}%
\bibitem [{\citenamefont {Gotoh}\ \emph {et~al.}(2002)\citenamefont {Gotoh},
  \citenamefont {Fukayama},\ and\ \citenamefont {Nakano}}]{Gotoh2002}%
  \BibitemOpen
  \bibfield  {author} {\bibinfo {author} {\bibfnamefont {T.}~\bibnamefont
  {Gotoh}}, \bibinfo {author} {\bibfnamefont {D.}~\bibnamefont {Fukayama}},\
  and\ \bibinfo {author} {\bibfnamefont {T.}~\bibnamefont {Nakano}},\
  }\bibfield  {title} {\bibinfo {title} {{Velocity field statistics in
  homogeneous steady turbulence obtained using a high-resolution direct
  numerical simulation}},\ }\href {https://doi.org/10.1063/1.1448296}
  {\bibfield  {journal} {\bibinfo  {journal} {Phys. Fluids}\ }\textbf {\bibinfo
  {volume} {14}},\ \bibinfo {pages} {1065} (\bibinfo {year}
  {2002})}\BibitemShut {NoStop}%
\bibitem [{\citenamefont {Ishihara}\ \emph {et~al.}(2007)\citenamefont
  {Ishihara}, \citenamefont {Kaneda}, \citenamefont {Yokokawa}, \citenamefont
  {Itakura},\ and\ \citenamefont {Uno}}]{Ishihara2007}%
  \BibitemOpen
  \bibfield  {author} {\bibinfo {author} {\bibfnamefont {T.}~\bibnamefont
  {Ishihara}}, \bibinfo {author} {\bibfnamefont {Y.}~\bibnamefont {Kaneda}},
  \bibinfo {author} {\bibfnamefont {M.}~\bibnamefont {Yokokawa}}, \bibinfo
  {author} {\bibfnamefont {K.}~\bibnamefont {Itakura}},\ and\ \bibinfo {author}
  {\bibfnamefont {A.}~\bibnamefont {Uno}},\ }\bibfield  {title} {\bibinfo
  {title} {{Small-scale statistics in high-resolution direct numerical
  simulation of turbulence: Reynolds number dependence of one-point velocity
  gradient statistics}},\ }\href {https://doi.org/10.1017/S0022112007008531}
  {\bibfield  {journal} {\bibinfo  {journal} {J. Fluid Mech.}\ }\textbf
  {\bibinfo {volume} {592}},\ \bibinfo {pages} {335} (\bibinfo {year}
  {2007})}\BibitemShut {NoStop}%
\bibitem [{\citenamefont {Ishihara}\ \emph {et~al.}(2009)\citenamefont
  {Ishihara}, \citenamefont {Gotoh},\ and\ \citenamefont
  {Kaneda}}]{Ishihara2009}%
  \BibitemOpen
  \bibfield  {author} {\bibinfo {author} {\bibfnamefont {T.}~\bibnamefont
  {Ishihara}}, \bibinfo {author} {\bibfnamefont {T.}~\bibnamefont {Gotoh}},\
  and\ \bibinfo {author} {\bibfnamefont {Y.}~\bibnamefont {Kaneda}},\
  }\bibfield  {title} {\bibinfo {title} {{Study of high-reynolds number
  isotropic turbulence by direct numerical simulation}},\ }\href
  {https://doi.org/10.1146/annurev.fluid.010908.165203} {\bibfield  {journal}
  {\bibinfo  {journal} {Annu. Rev. Fluid Mech.}\ }\textbf {\bibinfo {volume}
  {41}},\ \bibinfo {pages} {165} (\bibinfo {year} {2009})}\BibitemShut
  {NoStop}%
\bibitem [{\citenamefont {Stauffer}\ and\ \citenamefont
  {Amnon}(1994)}]{Stauffer1994}%
  \BibitemOpen
  \bibfield  {author} {\bibinfo {author} {\bibfnamefont {D.}~\bibnamefont
  {Stauffer}}\ and\ \bibinfo {author} {\bibfnamefont {A.}~\bibnamefont
  {Amnon}},\ }\href@noop {} {\emph {\bibinfo {title} {{Introduction to
  percolation theory: revised second edition}}}}\ (\bibinfo  {publisher}
  {General $\&$ Introductory Physics},\ \bibinfo {year} {1994})\BibitemShut
  {NoStop}%
\bibitem [{\citenamefont {Shrivastav}\ \emph {et~al.}(2016)\citenamefont
  {Shrivastav}, \citenamefont {Chaudhuri},\ and\ \citenamefont
  {Horbach}}]{Shrivastav2016}%
  \BibitemOpen
  \bibfield  {author} {\bibinfo {author} {\bibfnamefont {G.~P.}\ \bibnamefont
  {Shrivastav}}, \bibinfo {author} {\bibfnamefont {P.}~\bibnamefont
  {Chaudhuri}},\ and\ \bibinfo {author} {\bibfnamefont {J.}~\bibnamefont
  {Horbach}},\ }\bibfield  {title} {\bibinfo {title} {{Yielding of glass under
  shear: A directed percolation transition precedes shear-band formation}},\
  }\href {https://doi.org/10.1103/PhysRevE.94.042605} {\bibfield  {journal}
  {\bibinfo  {journal} {Phys. Rev. E}\ }\textbf {\bibinfo {volume} {94}},\
  \bibinfo {pages} {042605} (\bibinfo {year} {2016})}\BibitemShut {NoStop}%
\bibitem [{\citenamefont {Cao}\ \emph {et~al.}(2018)\citenamefont {Cao},
  \citenamefont {Dahmen}, \citenamefont {Kushima}, \citenamefont {Wright},
  \citenamefont {Park}, \citenamefont {Short},\ and\ \citenamefont
  {Yip}}]{Cao2018}%
  \BibitemOpen
  \bibfield  {author} {\bibinfo {author} {\bibfnamefont {P.}~\bibnamefont
  {Cao}}, \bibinfo {author} {\bibfnamefont {K.~A.}\ \bibnamefont {Dahmen}},
  \bibinfo {author} {\bibfnamefont {A.}~\bibnamefont {Kushima}}, \bibinfo
  {author} {\bibfnamefont {W.~J.}\ \bibnamefont {Wright}}, \bibinfo {author}
  {\bibfnamefont {H.~S.}\ \bibnamefont {Park}}, \bibinfo {author}
  {\bibfnamefont {M.~P.}\ \bibnamefont {Short}},\ and\ \bibinfo {author}
  {\bibfnamefont {S.}~\bibnamefont {Yip}},\ }\bibfield  {title} {\bibinfo
  {title} {{Nanomechanics of slip avalanches in amorphous plasticity}},\ }\href
  {https://doi.org/10.1016/j.jmps.2018.02.012} {\bibfield  {journal} {\bibinfo
  {journal} {J. Mech. Phys. Solids}\ }\textbf {\bibinfo {volume} {114}},\
  \bibinfo {pages} {158} (\bibinfo {year} {2018})}\BibitemShut {NoStop}%
\end{thebibliography}


%

\end{document}